\documentclass[12pt]{article}

\usepackage{graphicx,amscd,amsmath,amssymb,verbatim}
\usepackage[TS1,OT1,T1]{fontenc}
\usepackage{anysize}
\usepackage{array}
\usepackage[T1]{fontenc}
\usepackage{bm}
\usepackage{times}
\usepackage{courier}
\usepackage{color}
\usepackage{fancyvrb}
\usepackage{graphicx}
\usepackage{subfig}

\newcommand\ket[1]{\ensuremath{|#1\rangle}}
\newcommand\bra[1]{\ensuremath{\langle#1|}}

\newcommand\oprod[2]{\ensuremath{|#1\rangle\langle#2|}}

\fvset{gobble=6,frame=single,framesep=2mm,commandchars=\\\{\}}

\def\>{\rangle}
\def\<{\langle}

\begin{document}

\title{Measurement-Based Quantum Computing with Valence-Bond-Solids}
\author{Leong Chuan Kwek $^{a,b,c}$\footnote{cqtklc@nus.edu.sg}, Zhaohui Wei $^{a}$\footnote{cqtwz@nus.edu.sg} and Bei Zeng$^{d,e}$ \footnote{zengb@uoguelph.ca}\\
$^a$ Centre for Quantum Technologies and Department of Physics, \\
 National University of
Singapore,  2 Science Drive 3, Singapore 117542\\
$^b$ Institute of Advanced
Studies (IAS), \\ Nanyang Technological University, Singapore 639673 \\
$^c$ National Institute of Education, \\
Nanyang Technological
University, Singapore 637616\\
$^d$ University of
Guelph, Department of Mathematics and Statistics\\
Guelph, Ontario, Canada N1G 2W1\\
$^e$ Institute for Quantum Computing, University of Waterloo\\
Waterloo, Ontario, Canada N2L 3G1}
\date{\today}
\maketitle

\begin{abstract}
Measurement-based quantum computing (MBQC) is a model of quantum
computing that proceeds by sequential measurements of individual
spins in an entangled resource state. However, it remains a
challenge to produce efficiently such resource states. Would it be
possible to generate these states by simply cooling a quantum
many-body system to its ground state? Cluster states, the canonical
resource states for MBQC,
 do not occur naturally as unique ground states of physical systems. This inherent hurdle
 has led
to a significant effort to identify alternative resource states that
appear as ground states in spin lattices. Recently, some interesting
candidates have been identified with various valence-bond-solid
(VBS) states. In this review, we provide a pedagogical introduction
to recent progress regarding MBQC with VBS states as possible
resource states.  This study has led to an interesting
interdisciplinary research area at the interface of quantum
information science and condensed matter physics.
\end{abstract}

\section{Introduction}

%%\noindent** A brief history of quantum computers

Although Richard Feynman ~\cite{Fey82} mooted the idea of a quantum
computer nearly thirty years ago in 1982, serious research work into
quantum computing probably only appeared about fifteen years ago
when Peter Shor demonstrated that a quantum computer, if available,
could perform prime number factorization exponentially faster than a
conventional computer which we typically call "classical computer".
A quantum computer essentially exploits inherently quantum
mechanical properties like the linear superposition principle and
entanglement to improve computational speed and complexity.

In Feynman's original idea, a quantum computer, if constructed,
would be also an ideal machine for simulating naturally-occurring
many-body quantum problems. After all, nature is largely quantum
mechanical and the ideal platform for solving problems in nature
should be a quantum mechanical device. From a theoretical computer
science perspective, the rigorous foundation of  a quantum computing
device was laid,  in 1985, by David Deutsch at Oxford who cleverly
formulated a theoretical model called the quantum Turing
machine~\cite{Deu85}, which has been a fundamental lego block for
the study of a quantum computer.

%%\noindent** The circuit model
In 1989, Deutsch introduced another model of fundamental importance
for quantum computation: the quantum circuit model~\cite{Deu89}. Yao
then showed in 1993  that the circuit model is actually equivalent
to the quantum Turing machine in terms of their computational
power~\cite{Yao93}. As we have mentioned before, Shor published his
famous algorithm for efficient prime numbers factorization on a
quantum computer in 1994~\cite{Sho94}. Factoring is believed to be a
difficult problem to solve on a classical computer and it lies at
the heart of modern cryptographic technology: the
Rivest-Shamir-Adleman (RSA) scheme. Shor's theoretical insight, if
realized, thus poses an immediate threat to classical cryptographic
security.  Equipped with quantum computers, all security systems
currently in use based on the computational difficulty of large
prime number factorization would be rendered useless.

The basic entity in classical computing is the bit. A bit is usually
encoded in several ways, for instance, the voltage or current  in a
wire or the energy in a capacitor. In a quantum computer, the
information can also be encoded in different ways: for instance, the
polarization of a photon or the energy levels of a two-level atomic
or molecular system. Unlike classical bit, the information for a
two-level quantum system can be encoded in a superposition of the
two levels of the system. Such encoding is called a quantum bit or a
qubit. Thus, qubits form the basic building blocks for a quantum
computer.

Classical computation proceeds essentially through a series of wires
and gates. The quantum circuit model in which algorithms are
implemented through a series of quantum gates thus remains the most
promising paradigm for the large scale realization of a quantum
computer. For a quantum computer, the circuit model applies a
sequence of unitary operations on an initial quantum state
$\ket{\psi_i}$ of a system with $n$ qubits and evolves the quantum
system according to the Schroedinger equation. Although each single
gate usually acts non-trivially on only one or two qubits, the
concatenation of these gates proves to be sufficient for producing
any unitary operation $U$ on the $n$-qubit initial state
$\ket{\psi_i}$, a property known as ``universal'' computation. The
application of this sequence of gates transforms an initial state
$\ket{\psi_i}$ into a final state $\ket{\psi_f}=U\ket{\psi_i}$. The
final computational result is then extracted from appropriate
measurements on the final state.

%%\noindent** Models of quantum computing

The realization of the quantum circuit model in various physical
systems has been extensively studied. These approaches include
implementation through nuclear magnetic resonance ~\cite{Cor04}, the
manipulations of atoms or ions in ion traps ~\cite{Win04}, the
manipulation of neutral atom ~\cite{Cav04}, implementation with
cavity QED~\cite{Cha04}, the optical platform with linear or
nonlinear optical devices~\cite{KMH04}, manipulation of electrons or
atoms in the solid state ~\cite{ACD+04}, superconducting
devices~\cite{Orl04}, and a ``unique'' qubit scheme~\cite{LHH04}.
Regardless of the approach in the realization of large scale quantum
computing, DiVincenzo has elegantly summarized five important
criteria for the physical implementation of future quantum
computers~\cite{Div00}, and each one of these  criteria  has been
carefully examined
~\cite{Win04,Cav04,Cha04,KMH04,ACD+04,Orl04,LHH04,Div00}. To date
none of these systems is capable of realizing a large scale quantum
computer in the foreseeable future without any glitches: each system
presents uniquely its own set of challenges and problems.  And
indeed one important stumbling block is that many quantum gates
entangling two qubits cannot be implemented with high fidelity in
many of these systems~\cite{Roa04}. To overcome some of these
limitations, researchers have sought other paradigms of quantum
computation. Although these alternative paradigms are equivalent to
the quantum circuit model in terms of computational power, they are
very different in terms of real physical realization. Among these
other paradigms,  some of more promising ones are measurement-based
quantum computing~\cite{RB01}, topological quantum
computing~\cite{KL09} and adiabatic quantum computing~\cite{FJS+00}.

%%%%%%%%%%%%%%I have replacd till here...more ;ater.......%%%%%%%%%%
%%\noindent** Measurement-based quantum computing

In this review, we focus on the measurement-based quantum computing
(MBQC) model, first introduced by Raussendorf and Briegel in
2001~\cite{RB01}.  MBQC is a model of quantum computing that is
equivalent to the circuit model in terms of computational power,
given that it can simulate every single-qubit gate and two-qubit
gate in the circuit model in an efficient way, and it is thus
universal for quantum computing. However, the physical realization
of MBQC is very different from that of the circuit model. It starts
from a highly entangled multi-qubit state $\ket{\Psi_C}$, called a
cluster state. Sequential single-qubit measurements are then
performed on the cluster state to realize quantum computing. The
advantage of MBQC is that once $\ket{\Psi_C}$ is prepared, no
entangling gates are needed in process of the computation. Or one
may imagine that all the entanglement needed for a quantum
computation is already embodied in $\ket{\Psi_C}$. MBQC thus seems
very appealing for the implementation of quantum computing provided
that $\ket{\Psi_C}$ can be efficiently prepared initially and
maintained during the process of measurements.

%%\noindent** Resource states for MBQC with VBS

Perhaps the most natural and simplest means of preparing an
entangled quantum state is to search for a many-body system in which
the unique ground state is the required entangled resource.  With
correct cooling techniques, one should then be able to acquire the
ground state and hence the entangled resource state needed for MBQC.
Unfortunately, it turns out that $\ket{\Psi_C}$ does not naturally
occur as unique ground states of any naturally-occurring physical
systems. More precisely, it was shown by Nielsen that $\ket{\Psi_C}$
cannot even be a unique ground state of any Hamiltonian of qubits
involving at most two-particle interactions~\cite{Nie06}. It is then
highly desirable seeking for other many-body quantum states, with
which MBQC can be performed in a similar manner as is performed on
$\ket{\Psi_C}$, where only single-particle measurements are needed
to implement universal quantum computing. We call these alternative
states for MBQC ``resource states''. Moreover, we want these
resource states to naturally occur as unique ground states of some
lattice spin systems, that is, unique ground states of Hamiltonians
involving only two-particle nearest-neighbor interactions on a
lattice. In addition, one also wants the resource states to be
stable during the process of measurements, that is, to remain as a
ground state of the Hamiltonian after the particles are measured and
discarded. To satisfy this condition, it is sufficient that the
Hamiltonian is ``frustration-free'', that is, the ground state of
the Hamiltonian is also a ground state of each interaction term of
the Hamiltonian. Another desired property is that the Hamiltonian is
gapped, meaning that it has a constant energy gap regardless of the
system size (i.e., the number of particles), so the ground state can
be stable against thermal fluctuation.

In 2007, Gross and Eisert took the first step towards the
construction of such realistic resource states for MBQC~\cite{GE07}.
They introduced an interesting framework for producing resource
states which is closely related to the theoretical description of
valence-bond-solids (VBS). The gist of their proposal is that even
if the quantum computation is actually performed on a system of
qubits, the actual physical system is not just a system of particles
with information encoded in two different levels. These particles
could ideally  possess higher spins -  a feature which greatly
facilitates the search for realistic lattice Hamiltonians. Their
framework allows one to construct various resource states with
different spin values and on different types of lattices. For
resource states associated with realistic Hamiltonians, they found
Hamiltonians on a spin-$1$ chain, involving only nearest-neighbor
interactions, whose unique ground states are resource states for
MBQC. Moreover, these systems are also frustration-free and gapped.
Unfortunately, their scheme applies specifically  to
one-dimensional(1D) spin chains. Since a 1D spin chain can only
process quantum computation of a single qubit, their framework are
not universal. It was also not obvious how one could extend the
scheme to find any realistic two-dimensional(2D) resource states for
MBQC. In 2008, Brennen and Miyake showed that a famous
one-dimensional VBS state, the Affleck-Kennedy-Lieb-Tasaki (AKLT)
state on a spin-$1$ chain~\cite{AKLT87} with certain boundary
condition, can act as a resource state for MBQC~\cite{BM08}.
Although this is still a result for 1D chain, it sharpens the
connection between the studies of MBQC and VBS states~\cite{KX10}.

In 2009, Chen {\it et al.} took an important step towards the
construction of 2D realistic resource states for MBQC~\cite{CZG+09}.
They showed that a state $\ket{\Psi_{triC}}$ of a spin-$5/2$ system
on the honeycomb lattice, called the tri-Cluster state, is a unique
ground state of a frustration-free, gapped, two-body Hamiltonian
with only nearest-neighbor interactions. Although the form of the
actual spin Hamiltonian appears complicated, their method does open
up new possibilities for constructing other 2D realistic resource
states for MBQC.  In early 2010, Cai {\it et al.} then constructed a
realistic resource state of a spin-$3/2$ system on the 2D octagonal
lattice, called a quantum magnet, which is in turn based on proper
coupling of spin-$3/2$ quasi-AKLT chains.  In the same year, Chen
{\it et  al.} developed an interesting alternative viewpoint on the
universality of VBS states for MBQC, by showing that these known
resource states can be reduced to the cluster states via adaptive
local measurements at a constant cost~\cite{CDJ+10}. This viewpoint
turns out to be particularly useful for studying the universality of
VBS states for MBQC, as cluster states themselves can be viewed as
VBS states that was first realized by Verstraete and Cirac in
2004~\cite{VC04}. Also based on this viewpoint, Wei {\it et al.}
reinterpreted the universality of the quantum magnet by reducing
this resource states to a 2D cluster state~\cite{WRK11}.
Interestingly, Wei {\it et  al.} and Miyake independently discovered
that the 2D AKLT state on the honeycomb lattice can also be a
resource state for MBQC~\cite{WAR11,WAR10,Miy10}. Although it is
still unclear if the 2D AKLT Hamiltonian on the honeycomb lattice is
gapped~\cite{AKLT88}, this result nicely links the study of MBQC and
VBS.

The ``quick reduction'' from spin-$5/2$ to spin-$3/2$ systems for
realistic 2D resource states for MBQC naturally gave further impetus
for the search of realistic resource states in a spin-$1/2$ system.
Unfortunately, it was recently shown by Chen {\it et al.}  that this
may not be possible. Chen {\it et al.} provided a no-go theorem
showing that there does not exist a unique ground state for a
two-body frustration-free Hamiltonian that can simultaneously be a
resource state for MBQC~\cite{CCD+10} for a qubit system. It was
further shown by Ji {\it et al.} that indeed for two-level systems,
the structure of the ground-state space for any two-body
frustration-free Hamiltonian can be fully characterized~\cite{JWZ10}
and none of these states corresponds to a resource state for MBQC.
These results herald bad news for the practical realization of MBQC
based on VBS resource states since spin-$1/2$ systems appear to be
the most prevalent systems in nature. One could try to overcome this
situation by relaxing the requirement for ``frustration-free'' or
``uniqueness'' and resort to some other physical mechanisms, such as
topological protection~\cite{Miy11,BBM+11} or
perturbation~\cite{NLD+08,BR08,BO08}.  Finally it is worth noting
that it is still an open question whether one can find a realistic
spin-$1$ VBS resource state for MBQC with a two-body Hamiltonian
which is both frustration-free and gapped.

%%\noindent** Organization of the paper

This review provides a pedagogical introduction to recent progress
regarding MBQC with VBS states, which, we believe, is an interesting
interdisciplinary research area at the interface of quantum
information science and condensed matter physics. We assume no prior
background on quantum computing of the readers, but we do assume
basic knowledge of quantum physics, condensed matter physics and VBS
states. For readers who are not familiar with VBS states, we would
like to refer them to a recent review regarding entanglement in VBS
states, which is published in the 2010 volume of this
journal~\cite{KX10}. We will cover the following topics in this
review.
\begin{itemize}
\item A brief introduction to the circuit model of quantum computing.
\item An introduction to MBQC based on cluster states: what a cluster state is, how MBQC with cluster states simulates the quantum circuit model.
\item A viewpoint on the universality of the resource states for MBQC introduced in~\cite{CDJ+10}.
\item MBQC in 1D valence-bond chains, in particular, the universality of the 1D AKLT chain for processing a single-qubit information.
\item The spin-$5/2$ tri-cluster state $\ket{\Psi_{triC}}$ on the honeycomb lattice, its universality for MBQC and the properties of the corresponding tri-cluster Hamiltonian.
\item Resource states for MBQC in spin-$3/2$ systems, including the quantum magnets on the 2D octagonal lattice and the 2D AKLT state on the honeycomb lattice.
\item The no-go theorem of resource states for MBQC in spin-$1/2$ systems.
\end{itemize}

\section{The circuit model of quantum computing}
\label{sec:circuit}

The basis for a two-level quantum system (qubit) is typically
denoted by $\ket{0}$ and $\ket{1}$. A quantum operation on a qubit
is a $2\times 2$ unitary matrix, called a ``single-qubit quantum
gate''. The Pauli matrices
\begin{equation}
\label{eq:Pauli}
X=\begin{pmatrix} 0 & 1 \\1 & 0 \end{pmatrix},
\quad \text{and} \quad
Y=\begin{pmatrix} 0 & -i \\i & 0 \end{pmatrix},
\quad \text{and} \quad
Z=\begin{pmatrix} 1 & 0 \\0 & -1 \end{pmatrix},
\end{equation}
together with the identity opertator
\begin{equation}
\label{eq:id}
I=\begin{pmatrix} 1 & 0 \\0 & 1\end{pmatrix}
\end{equation}
form a basis for $2\times 2$ matrices.
The Hadamard gate $H$ is given by
\begin{equation}
\label{eq:H}
H=\frac{1}{\sqrt{2}}\begin{pmatrix} 1 & 1 \\1 & -1\end{pmatrix}.
\end{equation}
And other important single-qubit gates are the $X,Y,Z$ rotations given by
\begin{equation}
\label{eq:Pauli}
X_{\theta}=\exp(-i\theta X/2),
\quad \text{and} \quad
Y_{\theta}=\exp(-i\theta Y/2),
\quad \text{and} \quad
Z_{\theta}=\exp(-i\theta Z/2).
\end{equation}

A basis for an $n$-qubit system is chosen as the tensor products of
$\ket{0}$s and $\ket{1}$s. For instance, for $n=2$, the four basis
states are
\begin{equation}
\ket{0}\otimes\ket{0},\ \ket{0}\otimes\ket{1},\ \ket{1}\otimes\ket{0},\ \ket{1}\otimes\ket{1},
\end{equation}
which are in short written as
\begin{equation}
\label{eq:ba2}
\ket{00},\ \ket{01},\ \ket{10},\ \ket{11}.
\end{equation}

Unitary operations acting on two qubits are called ``two-qubit
quantum gates''.  The most-commonly used two-qubit quantum gate is the
controlled-NOT gate, which takes $\ket{x}\otimes\ket{y}$ to
$\ket{x}\otimes\ket{y\oplus x}$, where $x,y\in\{0,1\}$ and $\oplus$
is the addition $\mod\ 2$. Here the first qubit is called the
control qubit, which remains unchanged, and the second qubit is
called the target qubit, which is flipped if the control qubit is
$1$. In the basis of Eq.(\ref{eq:ba2}) the matrix of a
controlled-NOT gate is then given by
\begin{equation}
\label{eq:CNOT}
\begin{pmatrix} 1 & 0 & 0 & 0 \\0 & 1 & 0 & 0 \\ 0 & 0 & 0 & 1\\ 0 & 0 & 1 & 0 \end{pmatrix}.
\end{equation}
Similarly, a controlled-NOT gate with the second qubit the control qubit takes $\ket{x}\otimes\ket{y}$ to $\ket{x\oplus y}\otimes\ket{y}$.

Another important two-qubit gate is the controlled-$Z$ gate, which
transforms the basis of Eq.(\ref{eq:ba2}) in the following way:
\begin{equation}
\label{eq:cZ}
\ket{00}\rightarrow\ket{00},\ \ket{01}\rightarrow\ket{01},\ \ket{10}\rightarrow\ket{10},\ \ket{11}\rightarrow -\ket{11}.
\end{equation}
Given that the controlled-$Z$ gate is symmetric between the two qubits, it is not necessary to specify which one is the control qubit
and which one is the target qubit.

In the circuit model of quantum computing, the initial $n$-qubit state $\ket{\psi_i}$ is usually
chosen as the all $\ket{0}$ state $\ket{0}\otimes\ket{0}\cdots\otimes\ket{0}$, which is in short written as $\ket{00\cdots 0}$ or $\ket{0}^{\otimes n}$.
Then a sequence of single- and two-qubit quantum gates are applied on $\ket{\psi_i}$ to result in a final state $\ket{\psi_f}$. And finally single-qubit measurements are performed on each qubit, usually in the $\{\ket{0},\ket{1}\}$ basis, to obtain the result of the computation.

Indeed, if one is able to implement any two-qubit quantum gate on
any two qubits, then any $n$-qubit unitary operation can be
implemented by applying a sequence of two-qubit unitary
gates~\cite{NC00}. In short, all two-qubit unitary gates together
are universal for quantum computing, meaning that they can be used
to implement any quantum computation. Moreover, any two-qubit
unitary gate can be implemented through a sequence of single-qubit
unitary gates together with the controlled-NOT (or controlled-$Z$)
gate~\cite{NC00}.  Therefore, single-qubit gates plus the
controlled-NOT gate (single-qubit gates plus the controlled-$Z$
gate) are  also universal for quantum computing.

\begin{figure}[htbp]
  \centering
  \includegraphics[width=3in]{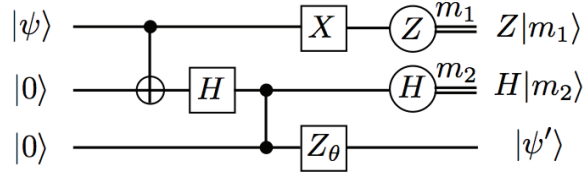}
  \caption{A quantum circuit}
\label{fig:Cir1}
\end{figure}

\begin{figure}[htbp]
  \centering
  \includegraphics[width=4in]{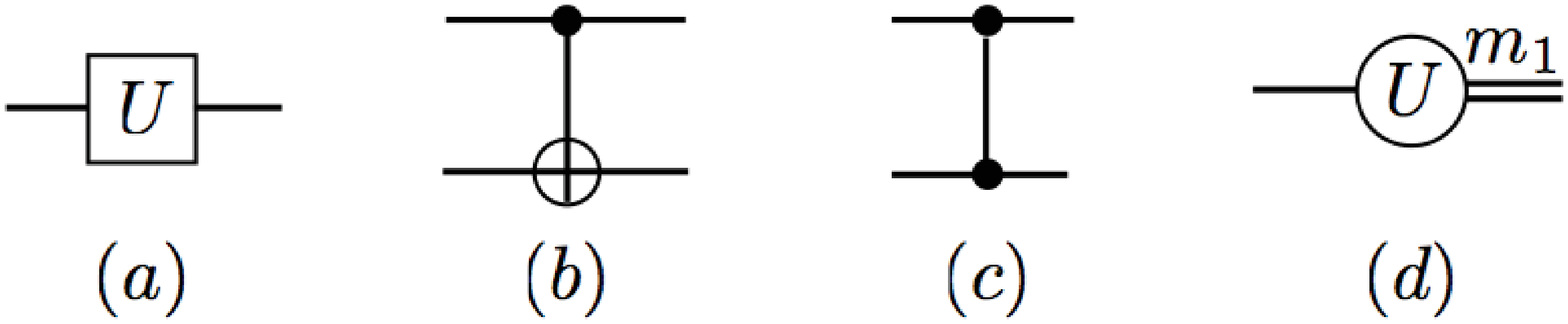}
  \caption{(a) A single-qubit gate $U$; (b) A controlled-NOT gate with the top qubit as the control qubit and the bottom qubit as the target qubit; (c) A controlled-Z gate; (d) A single-qubit measurement in the basis $\{U\ket{0},U\ket{1}\}$, and the measurement result $m_1$ is obtained which will give an output state $U\ket{m_1}$.}
\label{fig:Cir2}
\end{figure}

To implement a particular algorithm, a circuit diagram is typically
used to describe a quantum circuit. An example of a circuit diagram
is given in Fig.~\ref{fig:Cir1}. Each wire represents a qubit, and
the time goes from left to right. In this example, there are total
three qubits involved. The very left side gives the initial state of
the qubits, which is $\ket{\psi}\otimes\ket{0}\otimes\ket{0}$, where
$\ket{\psi}$ is a single-qubit state. The very right side gives the
final state of the qubits, which is $Z\ket{m_1}\otimes
H\ket{m_2}\otimes\ket{\psi'}$, where $m_1,m_2\in\{0,1\}$ and
$\ket{\psi'}$ is some single-qubit state whose actual value depends
on $\ket{\psi}$ and $\theta$. Each quantum gate in
Fig.~\ref{fig:Cir1} is explained in Fig.~\ref{fig:Cir2}.

\section{Measurement-based quantum computing}
\label{sec:MBQC}

\begin{figure}[htbp]
  \centering
  \includegraphics[width=3in]{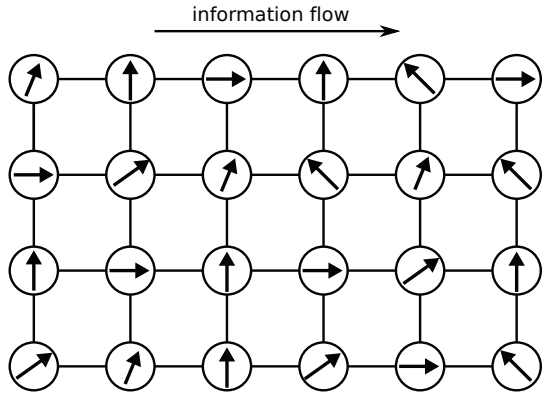}
  \caption{Measurement-Based Quantum Computing}
\label{fig:Clu2}
\end{figure}

This section provides an introduction to MBQC based on cluster
states, first introduced by Raussendorf and Briegel in
2001~\cite{RB01}. First let us briefly describe the way MBQC works.
Fig.~\ref{fig:Clu2} provides a schematic illustration of MBQC. In
this figure, each circle represents a qubit, which sits on a
two-dimensional square lattice. Each pair of qubits that are linked
by a solid line are called neighbors. The system of qubits is
prepared to an initial cluster state $\ket{\Psi_C}$. Then each qubit
will be measured sequentially, from the left columns to the right
columns, so the information flows from left to right. The arrow in
each circle illustrates the direction of the spin that is measured,
and the choice of directions are dependent on the results of earlier
measurements. At the end of the procedure, all the spins are
measured, and the measurement results together give the result of
the computation. There is another name used for MBQC in literature,
namely ``one-way quantum computing'', as the initial cluster state
is completely destroyed after the computation.

In Sec.~\ref{sec:cluster}, we will take a closer look at the cluster state. Then in Sec.~\ref{sec:gates}, we show how MBQC can simulate any quantum circuit in an efficient way, hence universal quantum computing can be implemented by MBQC.

\subsection{The cluster state}
\label{sec:cluster} 
\begin{figure}[htbp]
  \centering
  \includegraphics[width=3in]{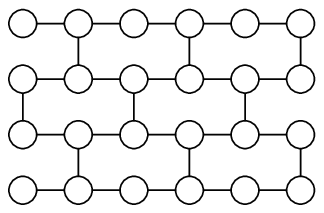}
  \caption{The honeycomb lattice}
\label{fig:Clu3}
\end{figure}

\begin{figure}[htbp]
  \centering
  \includegraphics[width=4in]{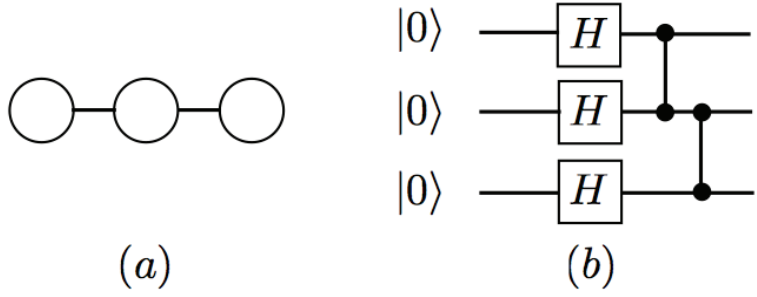}
  \caption{(a) A three-qubit graph; (b) A circuit creates the cluster state associated with the three-qubit graph}
\label{fig:Clu4}
\end{figure}

The cluster state was first introduced by
Briegel and Raussendorf shortly before they introduced the model of
MBQC~\cite{BR01}. The term ``cluster state'' actually refers to a
family of quantum states, which are quantum states of qubits
associated with certain lattices (or just simply arbitrary graphs).
For any graph of $n$ vertices, one can then define a cluster state.
For examples, the graph can be the 2D square lattice discussed in
Fig.~\ref{fig:Clu2}, the 2D honeycomb lattice given in
Fig.~\ref{fig:Clu3}, or a simple graph of three vertices given in
Fig. ~\ref{fig:Clu4}(a).

The creation of the cluster state associated with a given graph can
be described, in terms of quantum circuit, in three steps.
\begin{itemize}
\item Initialize every qubit in the state $\ket{0}$.
\item Apply a Hadamard gate on each of the qubits.
\item Apply a controlled-$Z$ gate on each pair of qubits who are neighbors on the graph (i.e., whose corresponding graph vertices are connected by a solid line).
\end{itemize}
As an example, a quantum circuit creates the cluster state
associated with the graph of three vertices given in Fig.
~\ref{fig:Clu4}(a) is given in Fig. ~\ref{fig:Clu4}(b).

For each vertex $j$ in a given graph, denote the neighboring qubits
of $j$ by $\text{nb}(j)$. Now we show that the cluster state
$\ket{\Psi_C}$ associated with the graph is an eigenstate of the
operator $X_{j}\bigotimes_{k\in\text{nb}(j)}Z_{k}$ with eigenvalue
$1$, where $X_j$ ($Z_k$) is the Pauli $X$ ($Z$) operator acting on
the $j$th ($k$th) qubit. That is,
\begin{equation}
\label{eq:eig1}
X_{j}\bigotimes_{k\in\text{nb}(j)}Z_{k}\ket{\Psi_C}=\ket{\Psi_C}
\end{equation}

We again take the simple example given in Fig. ~\ref{fig:Clu4}. We denote the qubits from left to right as $1,2,3$, and the controlled-$Z$ gate acting on the $j$th and $k$th qubits by $S_{jk}$. Then the cluster state associated with the graph is given by (according to the circuit given by \ref{fig:Clu4}(b))
\begin{equation}
\label{eq:3clus}
\ket{\Psi_C}=S_{23}S_{12}H_1H_2H_3\ket{000},
\end{equation}
where $H_j$ is the Hadamard operator acting on the $j$-th qubit. Now
we show that Eq.~(\ref{eq:eig1}) holds. That is,
\begin{eqnarray}
\label{eq:sta}
X_1Z_2\ket{\Psi_C}&=&\ket{\Psi_C},\nonumber\\
Z_1X_2Z_3\ket{\Psi_C}&=&\ket{\Psi_C},\nonumber\\
Z_2X_3\ket{\Psi_C}&=&\ket{\Psi_C}.
\end{eqnarray}
To show this, observe the following identities:
\begin{eqnarray}
\label{eq:com}
Z_jS_{jk}&=&S_{jk}Z_j\quad \text{and} \quad  X_jS_{jk}=S_{jk}X_jZ_k,\nonumber\\
Z_jH_j&=&H_jX_j\quad \text{and}\quad  X_jH_j=H_jZ_j.
\end{eqnarray}
Substituting Eq.~(\ref{eq:3clus}) into Eq.~(\ref{eq:com}) will
immediate result in Eq.~(\ref{eq:sta}). And it is straightforward to
show that Eq.~(\ref{eq:eig1}) holds for the cluster state associated
with any graph using a similar argument.

One can see that the operator $X_{j}\bigotimes_{k\in\text{nb}(j)}Z_{k}$ has eigenvalues $\pm 1$, and the operators corresponding to different vertices commute. For instance, $X_1Z_2$ and $Z_1X_2Z_3$ and $Z_2X_3$ commute with each other. Therefore, if we choose a Hamiltonian $H_C$ as
\begin{equation}
H_C=-\sum_jX_{j}\bigotimes_{k\in\text{nb}(j)}Z_{k},
\end{equation}
where the summation is running over all vertices of the graph, then
$\ket{\Psi_C}$ is obviously the unique ground state of the
Hamiltonian $H_C$. Moreover, this Hamiltonian is obviously gapped
(the entirely spectrum can be easily obtained) and frustration-free
(as $\ket{\Psi_C}$ is the ground state of each term in the
summation). Note that such a Hamiltonian is in general not a
two-body nearest-neighbor Hamiltonian.

\subsection{Simulation of basic quantum gates}
\label{sec:gates}

\begin{figure}[htbp]
  \centering
  \includegraphics[width=4.5in]{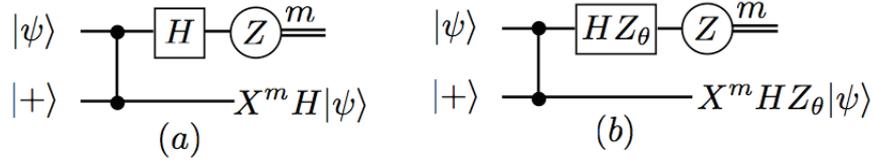}
  \caption{(This figure is redrawn from Eqs.(11) and (12) in~\cite{Nie06}). (a) A circuit for one-bit teleportation; (b) Generalized one-bit teleportation.}
\label{fig:Xtel}
\end{figure}

In this subsection, we show that MBQC based on certain cluster
states can simulate the quantum circuit model in an efficient way.
As discussed before, to implement universal quantum computing, one
only needs to realize single-qubit and controlled-NOT (or
controlled-$Z$) gates. Apart from the original discussion on how
MBQC simulates these gates~\cite{RB01}, there are other alternative
discussions~\cite{Nie03,Leu01,Leu03,AL04,CLN05,JP05,Nie06}. Here we
would like to take the one provided by Nielsen in~\cite{Nie06},
which is in turn based on the circuit given in
Fig.~\ref{fig:Xtel}(a) that is proposed in~\cite{ZLC00}.

The circuit in Fig.~\ref{fig:Xtel}(a) is known as one-bit teleportation, where $\ket{+}=\frac{1}{\sqrt{2}}(\ket{0}+\ket{1})$.
To show how it works, let $\ket{\psi}=\alpha\ket{0}+\beta\ket{1}$. Then note
\begin{eqnarray}
&&H_1S_{12}(\alpha\ket{0}+\beta\ket{1})\otimes \frac{1}{\sqrt{2}}(\ket{0}+\ket{1})\nonumber\\
&&=\frac{1}{\sqrt{2}}(\ket{0}\otimes H\ket{\psi}+\ket{1}\otimes XH\ket{\psi}).
\end{eqnarray}
Measuring the first qubit in the $\{0,1\}$ basis gives the desired
result. This circuit is then directly generalized to the one given
in Fig.~\ref{fig:Xtel}(b), as $Z_{\theta}$ commutes with $S_{12}$.

\begin{figure}[htbp]
  \centering
  \includegraphics[width=3.5in]{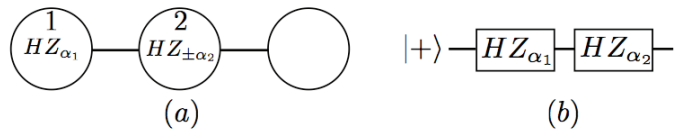}
  \caption{ (This figure is redrawn from Eqs.(13) and (14) in~\cite{Nie06}). (a) A graph of three vertices; (b) The single-qubit quantum circuit that can be simulated by MBQC associated with the graph in (a).}
\label{fig:Clu6}
\end{figure}

Now we show that MBQC associated with the simple graph of three
vertices given in Fig.~\ref{fig:Clu6}(a) simulates the circuit given
in Fig.~\ref{fig:Clu6}(b). Here $1,2$ in each circle of the graph
label the qubits $1$ and $2$. Then MBQC proceeds as follows: first
measure qubit $1$ in the basis of $HZ_{\alpha_1}\ket{m}$, where
$m=\{0,1\}$. If $m=0$ is obtained, then measure qubit $2$ in the
basis of $HZ_{\alpha_2}\ket{m'}$, where $m'=\{0,1\}$; if $m=1$ is
obtained, then measure qubit $2$ in the basis of
$HZ_{-\alpha_2}\ket{m'}$, where $m'=\{0,1\}$.

\begin{figure}[htbp]
  \centering
  \includegraphics[width=4.5in]{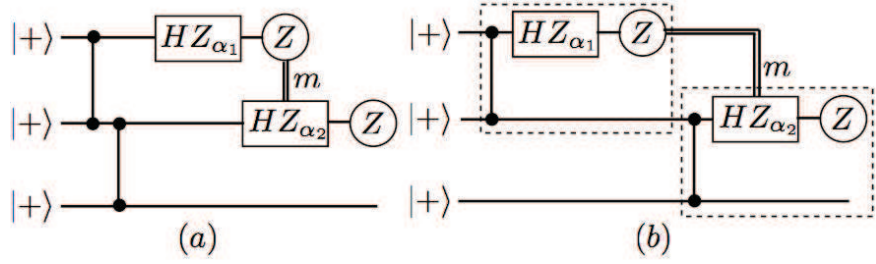}
  \caption{(This figure is redrawn from Eqs.(13) and (14) in~\cite{Nie06}). (a) The circuit corresponds to the measurement procedure in
  Fig.~\ref{fig:Xtel}(a); (b) The circuit equivalent to (a).}
\label{fig:Clu7}
\end{figure}

This procedure of MBQC can then be described by the quantum circuit given in Fig.~\ref{fig:Clu7}(a), which is equivalent to
the one in Fig.~\ref{fig:Clu7}(b). Now note here the circuit in each dashed box of Fig.~\ref{fig:Clu7}(b) is nothing
but the circuit of generalized one-bit teleportation given in Fig.~\ref{fig:Xtel}(b). It then follows that
MBQC associated with the simple graph given in Fig.~\ref{fig:Clu6}(a) simulates
the circuit given in Fig.~\ref{fig:Clu6}(b).

\begin{figure}[htbp]
  \centering
  \includegraphics[width=4in]{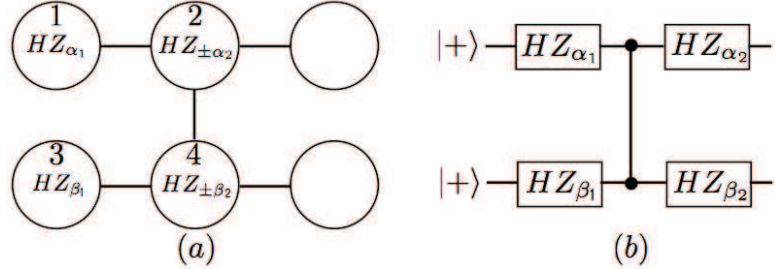}
  \caption{(This figure is redrawn from Eqs.(10) and (17) in~\cite{Nie06}). (a) A cluster state associated with a graph of six vertices; (b) The circuit corresponds to the measurement procedure in (a).}
\label{fig:Clu8}
\end{figure}

Note the ability to simulate the single-qubit gate of the form
$HZ_{\theta}$ for any $\theta$ suffices to perform any single-qubit
gate. This is because that $HHZ_{\theta}=Z_{\theta}$ and
$HZ_{\theta}H=X_{\theta}$, so both $X$ and $Z$ rotations of any
angle can be performed. And the idea for simulating the single-qubit
gates by MBQC also generalizes to two-qubit gates. For example, MBQC
associated with the graph given in Fig.~\ref{fig:Clu8}(a) simulates
the circuit of Fig.~\ref{fig:Clu8}(b). The proof proceeds  exactly
along the same line as the description for MBQC associated with
Fig.~\ref{fig:Clu6}(a), in which one can describe the procedure by a
similar quantum circuits given in Fig.~\ref{fig:Clu7}(a).

The ability to simulate the single-qubit gate of the form $HZ_{\theta}$ and the two-qubit gate given in Fig.~\ref{fig:Clu8}(b) thus
can implement the controlled-$Z$ gate. Together with arbitrary single-qubit gates, universal quantum computing can
then be simulated by MBQC.

\section{Resource states for MBQC}
\label{sec:resource}

It has been shown in Sec.~\ref{sec:gates} that single-qubit gates
and controlled-$Z$ gates can be simulated by MBQC based on cluster
states associated with certain kind of graphs. Universal quantum
computing can then be implemented by putting these graphs together,
which requires the graph to be of certain shape. Therefore not all
cluster states can be used as resource states for universal MBQC,
for instance cluster states associated with tree graphs (i.e., graph
without cycles) cannot be resource states for MBQC~\cite{SDV06}.
However, many of them are possible, for instance, the cluster states
associated with a square lattice or the honeycomb
lattice~\cite{VMD+06}.

The advantage of MBQC for the implementation of quantum computing in
the real world is that once the resource state is prepared, then
only single-qubit measurements are needed during the procedure of
computing. It then remains a challenge how to efficiently produce
such a resource state. To produce the cluster state, the direct way
is to use the three-step quantum circuit discussed in
Sec.~\ref{sec:cluster}. However, the third step requires applying
controlled-$Z$ gates on pairs of qubits. This then does not avoid
the difficulty of implementing entangling gates, meaning one does
not gain any advantage over the quantum circuit model.

An appealing idea to overcome entangling operations is to seek
physical systems in which one could obtain entangled resource states
as unique ground states and the generation of these ground states is
done through the cooling of quantum many-body systems. We already
know that the cluster state is a unique ground state of the
Hamiltonian $H_C$ given in Eq.~(\ref{eq:H}), and that $H_C$ is
gapped and frustration-free. However, $H_C$ involve many-body
interactions, not just two-body interactions and such Hamiltonians
are generally not easy to find in practice. Also, the $H_C$
associated with a 2D square lattice involves five-body interactions
and the $H_C$ associated with the 2D honeycomb lattice involves
four-body interactions. Even the $H_C$ associated with a 1D chain
involves at least three-body interactions.

One therefore hopes that there exists a Hamiltonian with only
two-body interactions that gives a particular cluster state as the
unique ground state. However, this turns out to be impossible as
observed by Nielsen~\cite{Nie06}. The idea behind the proof is the
following: if the Hamiltonian involves at most two-body
interactions, then its ground-state energy is totally determined by
the two-particle reduced density matrices of the ground states.
Indeed, for any cluster state $|\Psi_C\rangle$ that is a resource
state for MBQC, some of the eigenstates of $H_C$ have the same
two-particle reduced density matrices as $|\Psi_C\rangle$.
Therefore, if the cluster state $|\Psi_C\rangle$ is the ground
state, then some other eigenstates of $H_C$ are also ground states,
meaning the cluster state cannot be the unique ground state of any
two-body Hamiltonian.

As a result, it is then highly desirable to seek resource states
beyond the cluster states that are ``naturally-occurring''. We
consider MBQC with a general spin-$s$ system where $s$ might be
larger than $1/2$, meaning we do not restrict ourselves to qubit or
two-level systems. MBQC proceeds in a similar manner as in the qubit
system, but now sequential single-particle measurements are
performed on each spin-$s$ particle, each then with $2s+1$ possible
measurement outcomes. Of course, a spin-$1/2$ system is highly
desirable in practice, but simply by relaxing the restriction to
spin-$1/2$ systems provides us some new ideas in the search for
these highly entangled resource states. Accordingly, an ``ideal
state'' $\ket{\psi_{id}}$ associated with a Hamiltonian $H_{id}$
would satisfy the following conditions:
\begin{enumerate}
\item $\ket{\psi_{id}}$ is universal for MBQC, i.e., $\ket{\psi_{id}}$ is a resource state for MBQC.
\item $\ket{\psi_{id}}$ is the unique ground state of $H_{id}$.
\item $H_{id}$ is a lattice spin Hamiltonian with only nearest-neighbor two-body interactions.
\item $H_{id}$ has a constant energy gap.
\item $H_{id}$ is frustration-free, meaning that $\ket{\psi_{id}}$ is the ground state of each interaction term of $H_{id}$.
\end{enumerate}
Condition $1-4$ can be understood easily. Condition $5$ ensures the
stability of $\ket{\psi_{id}}$ throughout the process of MBQC, that
is, after some particles are measured and discarded, the state of
the unmeasured particles continues to remain in the ground-state
space of $H_{id}$. Even if all five conditions are met, we would
still like the value of spin $s$ to be as small as possible. At a
first glance, these demands appear daunting. Each condition alone is
difficult to fulfill in general. However, we now argue that there
are indeed simple examples which can reasonably satisfy all the
conditions given.

We begin with Condition $1$. To discuss what kind of spin states are
resource states for MBQC, one obvious way is to show that it can
simulate the quantum circuit model in an efficient way. In practice
this is not so easy to check. An important step was performed by
Gross and Eisert in 2007~\cite{GE07}. They introduced a framework
for producing resource states, that is closely related to the
theoretical description of valence-bond-solids (VBS) states and that
allows one to construct various resource states with different spin
values and on different type of lattices. Here in this review we
take another viewpoint, which is also practical and readily
applicable to VBS states. This main idea introduced by Chen {\it et
al.} ~\cite{CDJ+10} is the following observation:
\medskip

\noindent \textit{If a spin state $\ket{\psi}$ can be reduced to a
resource cluster state $\ket{\Psi_C}$ via adaptive local
measurements at a constant cost, then $\ket{\psi}$ is a resource
state for MBQC}.

\medskip
\noindent Moreover, this obviously sufficient condition for resource
states might also be necessary. As shown by Chen {\it et al.}, all
the known resource states do satisfy this condition. It turns out
that this sufficient condition is convenient to apply for checking
the universality for VBS states. To understand how this works, we
would first mention another interesting observation: the cluster
states can also be viewed as VBS states, which was first discussed
by Verstraete and Cirac in 2004~\cite{VC04}.

\begin{figure}[htbp]
  \centering
  \includegraphics[width=3.5in]{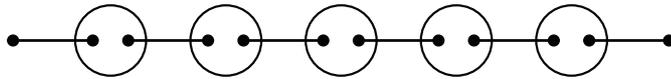}
  \caption{A VBS state on a chain.}
\label{fig:cluvbs1}
\end{figure}

We start from a simple example of the cluster state associated with
a 1D chain graph, as shown in Fig.~\ref{fig:cluvbs1}. Here each
circle represents a spin, called ``physical spin.'' And each black
dot also represents a spin, which we always choose as spin-$1/2$
throughout this review, hence is called ``virtual qubit.'' Each line
connecting two virtual qubits is called bond. The state of the two
virtual qubits connected by a bond is usually chosen as the singlet
state which is
\begin{equation}
\label{eq:singlet}
\ket{\psi_{singlet}}=\frac{1}{\sqrt{2}}(\ket{01}-\ket{10}).
\end{equation}

In the usual description of VBS states, there is a projection $P_j$
associated with the $j$-th physical spin, that projects the state of
two virtual qubits inside the circle onto a $d$-dimensional
subspace, hence the physical spin will be $d$-dimensional. For
instance, for the famous Affleck-Kennedy-Lieb-Tasaki (AKLT)
state~\cite{AKLT87}, each $P_j$ is chosen as the projection onto the
symmetric subspace of the two virtual qubits, resulting physical a
spin of dimension $3$, i.e., spin-$1$. And the quantum state of the
physical spins is then given by
\begin{equation}
\label{eq:1DVBS}
(\prod_j P_j)\bigotimes_{\alpha}\ket{\psi_{singlet}}_{\alpha},
\end{equation}
where the subscript $\alpha$ labels the singlets.

We now show that the cluster state associated with a chain graph
also possesses a VBS description, i.e., there exist projections
$P_j$'s that project the state of two virtual qubits inside each
circle onto a $2$-dimensional subspace, such that the state given by
Eq.(~\ref{eq:1DVBS}) is the 1D cluster state. . For convenience, we
choose the bond state as
\begin{equation}
\label{eq:clubond}
\ket{\psi_{cbond}}=\frac{1}{2}(\ket{00}+\ket{01}+\ket{10}-\ket{11})=H_1Z_1\otimes X_2\ket{\psi_{singlet}},
\end{equation}
which is equivalent to the singlet state by changing local basis of
the virtual qubits, as shown by the second equality.

Consider the projection is then chosen to be
\begin{equation}
\label{eq:cluP}
P=\ket{\tilde{0}}\bra{00}+\ket{\tilde{1}}\bra{11},
\end{equation}
which is the same for each physical qubit. Here
$\ket{\tilde{0}},\ket{\tilde{1}}$ are the basis of the physical
qubit. Recall that a cluster state associated with a line of $n$
vertices is given by
\begin{equation}
\ket{\Psi_C}=\prod_{j}S_{j,j+1}\prod_{j}H_j\ket{0}^{\otimes n}=\prod_{j}S_{j,j+1}\ket{+}^{\otimes n}.
\end{equation}
Note that
\begin{equation}
\ket{+}^{\otimes n}=\left(\frac{1}{\sqrt{2}}(\ket{0}+\ket{1})\right)^{\otimes n}=\frac{1}{2^{n/2}}\sum_{i_1,i_2,\ldots, i_n}\ket{i_1i_2\ldots i_n},
\end{equation}
where $i_l=\{0,1\}$. Therefore, for the term of $\prod_{j}S_{j,j+1}\ket{i_1i_2\ldots i_n}$, each string of the type $\ket{\ldots 11\ldots}$
contributes a $-1$ phase factor.

Now the VBS state of $n$ physical qubits with the projection $P$
given in Eq.(~\ref{eq:cluP}) as demonstrated in
Fig.~\ref{fig:cluvbs1}, starting from $n-1$ bonds given by
Eq.(~\ref{eq:clubond}), is given by
\begin{equation}\label{eq:1dresource}
\ket{\psi_V}=\prod_j P_j(\ket{\psi_{cbond}})^{\otimes n-1}.
\end{equation}
To understand the resulted state, one can expand
$(\ket{\psi_{cbond}})^{\otimes n-1}$ into a summation of terms of
the form $\ket{i_1i_2\ldots i_{2n-2}}$, with a phase factor either
$-1$ or $+1$. After the projections $\prod_j P_j$, only terms with
the strings $\ket{\ldots 00 \ldots}$ and $\ket{\ldots 11 \ldots}$
will remain. Here the two qubits of the state $00$ or $11$
correspond to any two virtual qubits in a same circle, which can
then be replaced by $\tilde{0}$ and $\tilde{1}$ after the
projections as the states of the physical qubits. Furthermore, given
the form of the bond state $\ket{\psi_{cbond}}$ as in
Eq.(\ref{eq:clubond}), each string of the type $\ket{\ldots
\tilde{1}\tilde{1}\ldots}$ contributes a $-1$ phase factor.
Therefore, the resulted state $\ket{\psi_V}$ is nothing but the
one-dimensional cluster state $\ket{\Psi_C}$.

Following a same argument, we choose the projections, instead, to be
\begin{equation}
\label{eq:cluP1}
P=\ket{\tilde{0}}\bra{01}+\ket{\tilde{1}}\bra{10},
\end{equation}
which is the same for each physical qubit, then the resulted state
$\ket{\psi_V}$ can also be the one-dimensional cluster state
$\ket{\Psi_C}$ after performing some local transformation which maps
$\ket{\tilde{0}}\rightarrow \ket{\tilde{1}}$ and
$\ket{\tilde{1}}\rightarrow \ket{\tilde{0}}$.

Let us now, consider, instead, the projection
\begin{equation}
\label{eq:cluP2}
P'=\ket{\tilde{0}}\bra{00}+\ket{\tilde{1}}\bra{11}+\ket{\tilde{2}}\bra{01}+\ket{\tilde{3}}\bra{10},
\end{equation}
which is the same for each physical qubit. Note this projection
actually projects each of the two virtual qubits in a same circle
onto the full $4$-dimensional Hilbert space. In other words, this
projection does nothing but a relabeling of the four basis. Denote
the resulted VBS state by $\ket{\psi'_V}$, which is actually a state
of a spin-$3/2$ system.

Now consider the measurement on each of the spins of
$\ket{\psi'_V}$, which is the projection onto either the
$\{\ket{\tilde{0}},\ket{\tilde{1}}\}$ or the
$\{\ket{\tilde{2}},\ket{\tilde{3}}\}$ subspace. Then according to
the discussions above, the resulted state is a state of $n$ qubits,
which can be transformed to the 1D cluster state $\ket{\Psi_C}$
after performing some local basis transformation dependent on each
of the measurement results. In this way,  $\ket{\psi'_V}$ reduces to
a 1D cluster state $\ket{\Psi_C}$ via adaptive local measurements at
a constant cost, where one copy of $\ket{\psi'_V}$ will give one
copy of $\ket{\Psi_C}$. Therefore, we know that $\ket{\psi'_V}$ is a
resource state for MBQC of a single qubit.

$\ket{\psi'_V}$ thus provides an example of the sort of states
described in~\cite{CDJ+10} regarding the usefulness of a quantum
state for MBQC. The advantage of using $\ket{\psi'_V}$ instead of
$\ket{\Psi_C}$ for MBQC of a single qubit is that $\ket{\psi'_V}$
can be associated with a Hamiltonian that satisfies all the
Conditions $2-5$. This is obvious, as $\ket{\psi'_V}$ is nothing but
bunch of singlets by viewing each pair of the virtual qubits in a
same circle as a spin-$3/2$ particle, whose corresponding
Hamiltonian involves only two-body nearest-neighbor interactions,
with $\ket{\psi'_V}$ the unique ground state, and is gapped and
frustration-free.

\begin{figure}[htbp]
  \centering
  \includegraphics[width=3.5in]{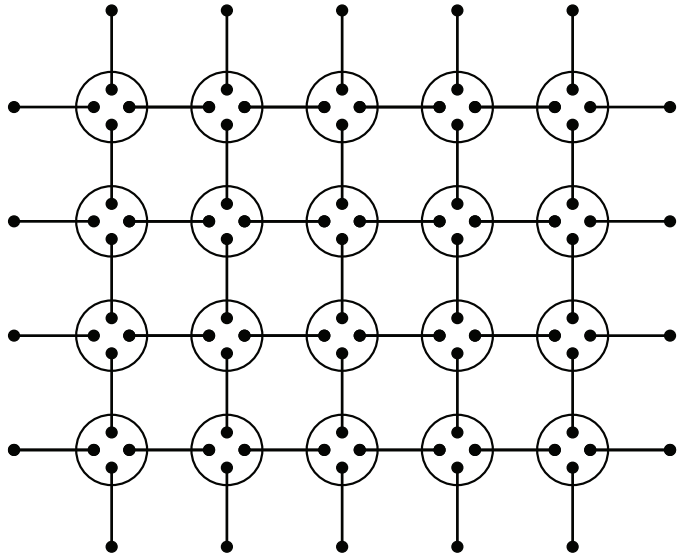}
  \caption{A VBS state on a square lattice}
\label{fig:cluvbs2}
\end{figure}

To generalize the above argument to 2D spin systems is just
straightforward. For the cluster state associated with a 2D square
lattice, as shown in Fig.~\ref{fig:cluvbs2}, one will just choose
the projection of each four virtual qubits inside a same circle as
\begin{equation}
\label{eq:cluP3}
P=\ket{\tilde{0}}\bra{0000}+\ket{\tilde{1}}\bra{1111}.
\end{equation}
Similarly, the VBS state by simply viewing each four of the virtual
qubits in a same circle as a $16$-dimensional particle is a resource
state for MBQC, which can be associated with a Hamiltonian that
satisfies the Conditions $2-5$.

\begin{figure}[htbp]
  \centering
  \includegraphics[width=3.5in]{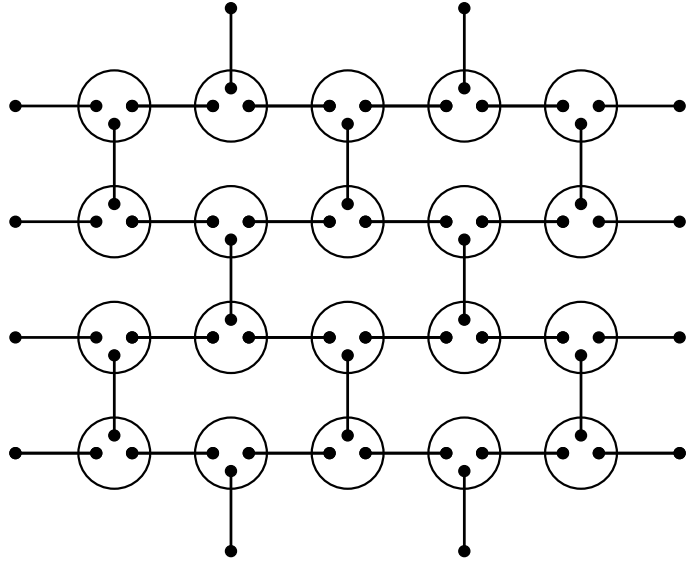}
  \caption{A VBS state on the honeycomb lattice}
\label{fig:cluvbs3}
\end{figure}

One can also consider the 2D honeycomb lattice. For the cluster
state associated with the 2D honeycomb lattice, as shown in
Fig.~\ref{fig:cluvbs3}, one will just choose the projection of each
three virtual qubits inside a same circle as
\begin{equation}
\label{eq:cluP3}
P=\ket{\tilde{0}}\bra{000}+\ket{\tilde{1}}\bra{111}.
\end{equation}
Similarly, the VBS state by simply viewing each three of the virtual
qubits in a same circle as a $8$-dimensional particle (i.e.,
spin-$7/2$) is a resource state for MBQC, which can be associated
with a Hamiltonian that satisfies Conditions $2-5$.

We now have successfully constructed simple examples of ``ideal
states'' which satisfy Conditions $1-5$ as desired. The drawback of
these constructions is obvious: each spin is of dimension $8$ to
perform universal MBQC. Even in the case for MBQC of a single qubit,
one will need a spin-$3/2$ system. This is not practical for the
realization of MBQC. So the most desirable thing is to reduce the
spin value but still keep Conditions $1-5$ satisfied. Compared to
the simple construction of ``ideal states'' provided in this
section, states with smaller spins turn out to be more difficult to
construct, which requires many more other techniques apart from the
sufficient condition for universality discussed above. This will be
the topic for the rest of this review. We will start from the 1D
situation in the next section.

\section{MBQC in 1D valence-bond chains}
\label{sec:1D}

In this section we consider 1D VBS states as described in
Fig.~\ref{fig:cluvbs1}. For convenience we would like to choose the
bond state to be
\begin{equation}
\label{eq:bond}
\ket{\psi_{bond}}=\frac{1}{\sqrt{2}}(\ket{00}+\ket{11})=Z_1\otimes X_2\ket{\psi_{singlet}},
\end{equation}
which is equivalent to $\ket{\psi_{singlet}}$ and
$\ket{\psi_{cbond}}$ by changing local basis. The advantage of
writing the bond state this way is the following. For each spin of
the system corresponding to two virtual qubits inside the circle,
let us write the projection as
\begin{equation}
P=\sum_m\ket{\tilde{m}}\sum_{k,l}A_{kl}[m]\bra{kl},
\end{equation}
where $k,l=\{0,1\}$. We can then view $A[m]$ as a $2\times 2$ matrix with entries $A_{kl}[m]$. And again we assume that the projection is the same for each physical spin, which is
reasonable in practice as usual physical systems do have this kind of translational symmetry.
The state
of the physical spins
\begin{equation}
\label{eq:bond}
\ket{\Psi}=(\prod_j P_j)\bigotimes_{\alpha}\ket{\psi_{bond}}_{\alpha},
\end{equation}
where the subscript $\alpha$ labels the bonds, will then have the following form
\begin{equation}
  \label{eq:mps}
  \ket{\Psi} = \sum_{i_1,\cdots,i_n} \bra{R}A[i_n]\cdots
  A[i_1]\ket{L} \ket{i_1\cdots i_n}.
\end{equation}
Here $i_j=0,1$, and $\ket{L},\ket{R}$ are left and write boundary conditions corresponding to the most left and most right virtual qubits in Fig.~\ref{fig:cluvbs1}.

As an example, for the 1D cluster state with a projection
$\ket{\tilde{0}}\bra{00}+\ket{\tilde{1}}\bra{11}$, which is
associated with the bond states $\ket{\psi_{cbond}}$, when writing in the form of Eq.(~\ref{eq:mps}), one can
equivalently choose the projection as
\begin{equation}
\label{eq:clumps}
A[0]=H,\quad\text{and}\quad A[1]=HZ.
\end{equation}
The state given by Eq.(\ref{eq:mps}) is called a matrix product
state (MPS). And it is known that all the 1D VBS states can be
written in this MPS form (for more discussions on MPS,
see~\cite{PVW+07}). The most famous VBS state is the
Affleck-Kennedy-Lieb-Tasaki (AKLT) state, which is the state of a
spin-$1$ chain~\cite{AKLT87}. The AKLT Hamiltonian of a spin-$1$
chain is
\begin{equation}
\label{eq:HAKLT}
H^{AKLT}=\sum_j\vec{S}_j\cdot\vec{S}_{j+1}+\frac{1}{3}(\vec{S}_j\cdot\vec{S}_{j+1})^2=\sum_{j}P_{j,j+1}^{(J=2)}-\frac{2}{3}.
\end{equation}
Here $\vec{S}_j$ is the spin operator of the $j$-th qubit, and
$P_{j,j+1}^{(J=2)}$ is the projection onto the total spin $J=2$
subspace of each neighboring pair of particles.

This Hamiltonian is known to be frustration-free and
gapped~\cite{AKLT87}. If we consider a finite chain of $n$ spin-$1$
particles, then the ground state is four-fold degenerate. One can
pick up a unique state by appending a spin-$1/2$ particle to each
end of the chain, and require that the total spin of the end
spin-$1/2$ particle and its neighboring spin-$1$ particle is $1/2$.
Denote this state by $\ket{\psi}_{AKLT}$, then the corresponding
Hamiltonian of which $\ket{\psi}_{AKLT}$ is a unique ground state is
\begin{equation}
\label{eq:HAKLTn}
H^{AKLT}_n=\sum_{j=1}^n\vec{S}_j\cdot\vec{S}_{j+1}+\frac{1}{3}(\vec{S}_j\cdot\vec{S}_{j+1})^2+\vec{s}_0\cdot\vec{S}_1+\vec{S}_n\cdot\vec{s}_{n+1},
\end{equation}
and $H^{AKLT}_n$ is also known to be frustration-free and gapped~\cite{AKLT88}.

The unique ground state of $H^{AKLT}_n$, i.e., the AKLT state
$\ket{\psi}_{AKLT}$, has a nice VBS representation~\cite{AKLT87}, as
is already mentioned in Sec.~\ref{sec:resource}. When choosing the
bond state as $\ket{\psi_{singlet}}$, the projection $P_j$
associated with each circle is to project onto the triplet subspace
(e.g the symmetric subspace) of two qubits, spanned by
\begin{equation}
\frac{1}{\sqrt{2}}(\ket{00}+\ket{11}),\quad\text{and}\quad \frac{1}{\sqrt{2}}(\ket{00}-\ket{11}),\quad\text{and}\quad \frac{1}{\sqrt{2}}(\ket{01}+\ket{10}).
\end{equation}
Associated with the bond states $\ket{\psi_{bond}}$, the projection
matrices based on the MPS representation are given by
\begin{equation}
\label{eq:akltmps} A[0]=X,\quad\text{and}\quad
A[1]=Y,\quad\text{and}\quad A[2]=Z.
\end{equation}

As mentioned before, it is already known that the AKLT state
$\ket{\psi}_{AKLT}$ satisfies Conditions $2-5$ in
Sec.~\ref{sec:resource}. Interestingly, it is shown by Brennen and
Miyake ~\cite{AKLT87} that $\ket{\psi}_{AKLT}$ is a resource state
for MBQC of a single qubit~\cite{BM08}. Their approach is based on
the earlier work of Gross and Eisert~\cite{GE07}, where they
developed a framework that shows how an MPS state (and its 2D
generalization) can simulate single-qubit (and two-qubit) gates via
single-particle measurements. This framework allows them to find
families of MPS states which satisfy Conditions $2-5$ given in
Sec.~\ref{sec:resource} and are resource states for MBQC of a single
qubit. These results are very nice progress toward finding practical
resource states for MBQC, which satisfy all of Conditions $1-5$
given in Sec.~\ref{sec:resource}. In this review we adopt another
approach that is discussed in Sec.~\ref{sec:resource}, i.e., the
sufficient condition of reducing the MPS state to a 1D cluster state
via adaptive local measurements at a constant cost.

In~\cite{CDJ+10}, a tabular form of MPS is introduced, which is
convenient for showing the reduction of other MPS states, including
the AKLT state, to a 1D cluster state with matrices given by
Eq.(\ref{eq:clumps}). In this tabular form, one writes the matrices
associated with each physical spin explicitly in a table. Here each
column of the table consists of the $2s+1$ matrices of a
corresponding physical spin of spin $s$. The physical indices
$\tilde{m}$'s determine a selection of the matrices $A[m]$ from each
column, whose product gives the correct amplitude together with the
boundary conditions $\bra{R}$ and $\ket{L}$, as given in
Eq.(\ref{eq:mps}). One then observes from Eq.~\eqref{eq:mps} that
the following properties hold~\cite{CDJ+10}:
\begin{enumerate}
\item For any two neighboring columns, multiplication of $M$ to the right of
all matrices in the left column and $M^{-1}$ to the left of all
matrices in the right column simultaneously does not change the state. This
can be directly seen from the form of Eq.~\eqref{eq:mps}, as
\begin{equation}
A[i_n]\cdots A[i_k]A[i_{k-1}] \cdots A[i_1]=A[i_n]\cdots A[i_k]MM^{-1}A[i_{k-1}] \cdots A[i_1].
\end{equation}
\item A unitary transformation in the physical space corresponds to
linear combinations of entries in the column with coefficients of the
unitary. This is a unitary transformation $U$ on the basis $\ket{\tilde{m}}$
that results in a new local basis $U\ket{\tilde{m}}$.
\item
The measurement in the computational basis corresponds to the
deletion of column entries not consistent with the measurement
outcome. This can be directly seen from the form of
Eq.~\eqref{eq:mps}.
\item
Columns of a single entry can be removed by absorbing them to a
neighboring column. This can also be directly seen from the form of
Eq.~\eqref{eq:mps}.
\end{enumerate}

As an example, Table 1 of Fig.~\ref{fig:IXZ} is a tabular form that
corresponds to two physical spins of the AKLT state, which consists
of two columns. Starting form this tabular form, we now show the
reduction from the AKLT state to the 1D cluster state, as discussed
in~\cite{CDJ+10}. We proceed to Table 2 of Fig.~\ref{fig:IXZ}, which
is obtained by adding the Y's with blue colour that represents the
same state as Table 1 according to property 1 of the tabular form.
Table 1 then gives the same state as Table 3 of Fig.~\ref{fig:IXZ},
up to local unitary transformations on the corresponding physical
basis, according to property 2 of the tabular form. Here in
Fig.~\ref{fig:IXZ}, $\simeq$ refers to equality up to local unitary
transformations. As a result, the AKLT state can also be represented
by the matrices $(I,X,Z)$ as in its MPS representation.

\begin{figure}[htbp]
  \centering
  \begin{tabular}[c]{m{3.1em} c m{4.3em} c m{3.1em}}
    \begin{Verbatim}[label=1]
      X X
      Y Y
      Z Z
    \end{Verbatim}
    & \parbox{1.5em}{\vspace{-.8em}$=$} &
    \begin{Verbatim}[label=2]
      X\textcolor{blue}{Y} \textcolor{blue}{Y}X
      Y\textcolor{blue}{Y} \textcolor{blue}{Y}Y
      Z\textcolor{blue}{Y} \textcolor{blue}{Y}Z
    \end{Verbatim}
    & \parbox{1.5em}{\vspace{-.8em}$\simeq$} &
    \begin{Verbatim}[label=3]
      Z Z
      I I
      X X
    \end{Verbatim}
  \end{tabular}\vspace{-1.5em}
  \caption{ (This figure is taken from FIG.1 in~\cite{CDJ+10}.) The MPS representation of the AKLT state by the matrices $(I,X,Z)$. }
  \label{fig:IXZ}
\end{figure}

Starting with this $(I,X,Z)$ form of the AKLT state, we perform two
different measurements $\mathcal{M}_1$ and $\mathcal{M}_2$
alternatively. Here $\mathcal{M}_1$ measures $\{\ket{\tilde{0}},
\ket{\tilde{1}}\}$ versus $\ket{\tilde{2}}$, and $\mathcal{M}_2$
measures $\{\ket{\tilde{0}}, \ket{\tilde{2}}\}$ versus
$\ket{\tilde{1}}$. That is, each measurement consists of a
two-dimensional projection and a one-dimensional projection. The
measurement outcomes are called success (failure) if the outcomes
correspond to the two(one)-dimensional subspaces, respectively. We
measure the two measurements sequentially along the AKLT chain, from
left to right, and switch the measurement we use only when the
previous one succeeds.

Table 1 in Fig.~\ref{fig:AKLT2Cluster} denotes a possible result
after these measurements on the AKLT state. More specifically, one
first measures $\mathcal{M}_1$ and succeeds. Next, the measurement
$\mathcal{M}_2$ is used. It fails and results in the
single-dimensional space $\ket{1}$, and we perform it again and it
succeeds subsequently, etc. After renaming the physical indices and
absorbing the $X$ and $Z$ in red color to their previous columns, we
will get Table 2 in Fig.~\ref{fig:AKLT2Cluster} by properties 4 and
2 of the tabular form. This then gives a 1D cluster state by the
second line of reasoning in Fig.~\ref{fig:AKLT2Cluster}, according
to property 2 of the tabular form.

\begin{figure}[htbp]
  \centering
  \begin{tabular}[c]{m{8.1em} c m{5.6em}}
    \begin{Verbatim}[label=1]
      I   I   I I
      X \textcolor{red}{X}     X
          Z \textcolor{red}{Z}   Z
    \end{Verbatim}
    & \parbox{1.5em}{\vspace{-.8em}$\simeq$} &
    \begin{Verbatim}[label=2]
      I I I I
      X Z X Z
    \end{Verbatim}
  \end{tabular}\\
  \begin{tabular}[c]{m{3.1em} c m{4.3em} c m{4.3em}}
    \begin{Verbatim}[label=3]
      I I
      X Z
    \end{Verbatim}
    & \parbox{1.5em}{\vspace{-.8em}$=$} &
    \begin{Verbatim}[label=4]
      I\textcolor{blue}{H} \textcolor{blue}{H}I
      X\textcolor{blue}{H} \textcolor{blue}{H}Z
    \end{Verbatim}
    & \parbox{1.5em}{\vspace{-.8em}$=$} &
    \begin{Verbatim}[label=5]
      H  H
      HZ HZ
    \end{Verbatim}
  \end{tabular}\vspace{-1.5em}
  \caption{(This figure is taken from FIG.2 in~\cite{CDJ+10}.) The reduction of the AKLT state to the 1D cluster state.}
  \label{fig:AKLT2Cluster}
\end{figure}

The above analysis based on the tabular can be directly generalized
to analyze other 1D resource states. For instance, the modified AKLT
state introduced in~\cite{GE07}, which is an MPS with
\begin{equation*}
  A[0] = H, \quad\text{and}\quad A[1]= X, \quad\text{and}\quad A[2] =Y,
\end{equation*}
can be similarly shown to be a resource state for MBQC of a single qubit.
And the one-parameter deformation of the AKLT model
considered by Fannes, Nachtergaele and Werner in Ref.~\cite{FNW92},
whose ground state
is an MPS with
\begin{equation*}
  A[0] = \sin\theta Z, \quad\text{and}\quad A[1]= \cos\theta \oprod{0}{1}, \quad\text{and}\quad A[2] =
  \cos\theta \oprod{1}{0},
\end{equation*}
can also be similarly shown to be a resource state for MBQC of a single qubit~\cite{CDJ+10}.

However, among all these 1D resource states for MBQC of a single
qubit, the AKLT state is of course the most interesting due to its
importance in the history of VBS. An experiment simulating MBQC on
an AKLT state, was performed in an optical system~\cite{KLZ+10}. In
this experiment, an AKLT state with a single site of spin-$1$
particle and two end spin-$1/2$ particles were prepared, and
rotations of the initial single-qubit state along any of the $X,Y,Z$
axis via MBQC were demonstrated.

\section{A resource state in a spin-$5/2$ system}
\label{sec:tricluster} In Sec.~\ref{sec:1D}, we discussed 1D
resource states which satisfy the natural Conditions $1-5$ given in
Sec.~\ref{sec:resource}. Note that the state $\ket{\psi'_V}$ given
by the projection of Eq.(\ref{eq:cluP2}) is a simple example for a
1D resource state, which is a spin-$3/2$ state. The advantage of the
states discussed in Sec.~\ref{sec:1D}, including the AKLT state, is
that they are spin-$1$ states.

To implement universal MBQC, we know that one will need some 2D
resource states which ideally satisfy the natural Conditions $1-5$
given in Sec.~\ref{sec:resource}. The spin-$7/2$ state on the
honeycomb lattice discussed in Sec.~\ref{sec:resource} provides such
an example, but with each physical particle of dimension $8$. One
would wish to reduce the spin dimension, but still keep Conditions
$1-5$ satisfied. This turns out to be a hard task, as although
states satisfying some of the conditions may be easy to find, it is
in general hard to show that they also satisfy the others.

The first important step of reducing the dimension is taken by Chen
{\it et al.} in~\cite{CZG+09}. They constructed a spin-$5/2$ VBS
state on the 2D honeycomb lattice. This 2D honeycomb lattice is
shown in Fig.~\ref{fig:cluvbs3}, where each bond state is chosen as
$\ket{\psi_{cbond}}$ as given in Eq.(\ref{eq:clubond}). Note that a
2D VBS state is also called a projective entanglement pair state
(PEPS)~\cite{VWP+06}. For each spin-$5/2$ associated with each
circle (with three virtual qubits inside), the corresponding
projection onto the six-dimensional subspace of the three-qubit
Hilbert space is chosen as
\begin{equation}
P_{triC}=\ket{\tilde{0}}\bra{000}+\ket{\tilde{1}}\bra{111}+\ket{\tilde{2}}\bra{100}+\ket{\tilde{3}}\bra{011}
+\ket{\tilde{4}}\bra{010}+\ket{\tilde{5}}\bra{101},
\end{equation}
and the corresponding PEPS state is called the tri-Cluster state, denoted by $\ket{\Psi_{triC}}$.

Based on the discussion in Sec.~\ref{sec:resource},
$\ket{\Psi_{triC}}$ is a resource state for MBQC, as it can be
reduced to the cluster state associated with the honeycomb lattice via
local measurements at a constant cost. More precisely,
$\ket{\Psi_{triC}}$ projected onto the subspace spanned by
$\{\ket{\tilde{0}}, \ket{\tilde{1}}\}$ is the same as the cluster
state, so are also the states given by $\ket{\Psi_{triC}}$ projected
onto $\{\ket{\tilde{2}},\ket{\tilde{3}}\}$ and
$\{\ket{\tilde{4}},\ket{\tilde{5}}\}$, up to local Pauli operations.

Now the task remains is to show that $\ket{\Psi_{triC}}$ also
satisfies Conditions $2-5$ in Sec.~\ref{sec:resource}. That is, one
would need to find a Hamiltonian involving only two-body
nearest-neighbor interactions, which is frustration-free, gapped,
and has $\ket{\Psi_{triC}}$ as its unique ground state. We start to
construct a frustration-free Hamiltonian $H_{triC}$ involving only
two-body nearest-neighbor interactions, which has
$\ket{\Psi_{triC}}$ as its ground state, and then further show that
the ground state is unique and $H_{triC}$ gapped.

\begin{figure}
  \centering
  \subfloat[]{\label{fig:peps2b1}\includegraphics[width=0.3\textwidth]{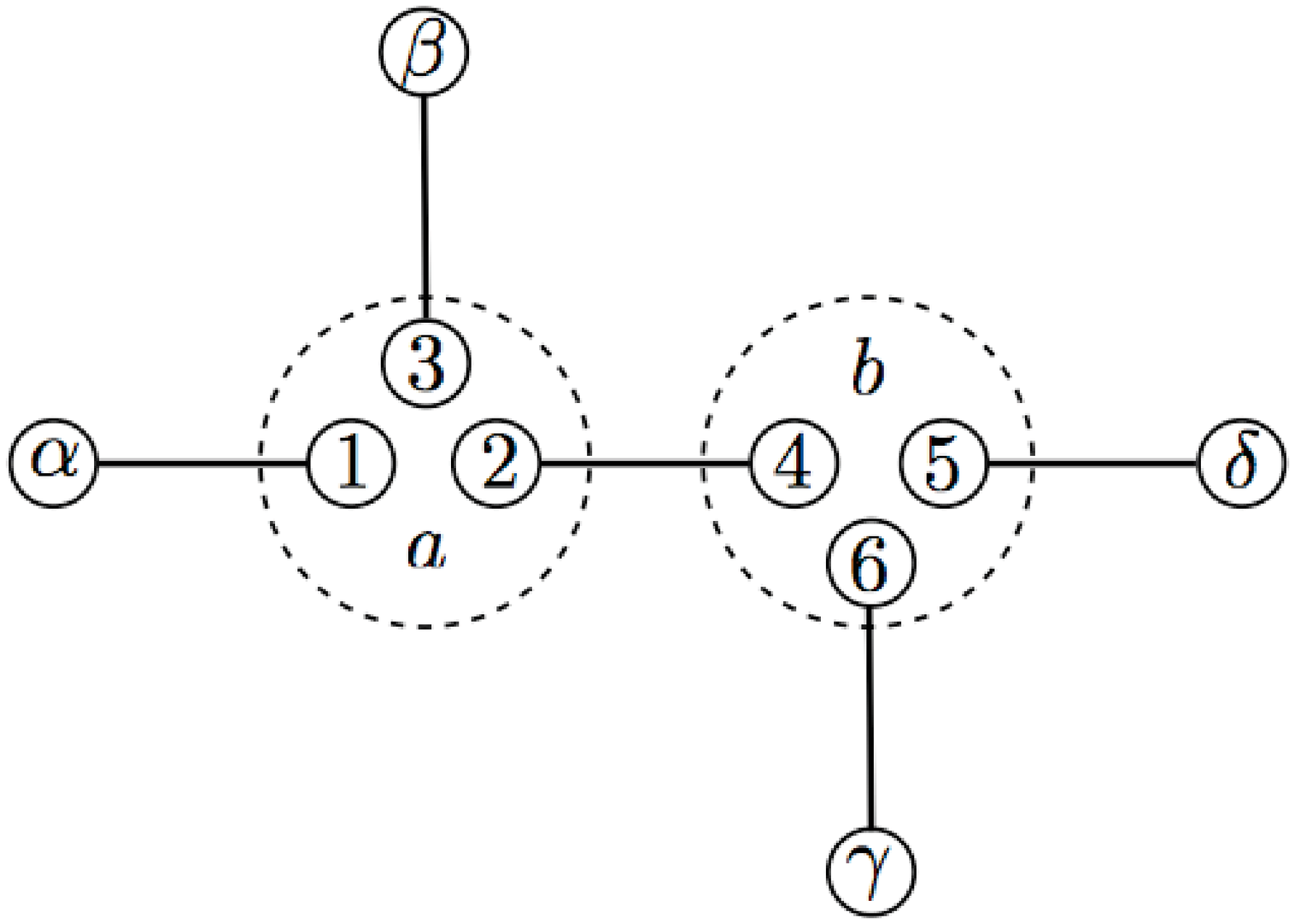}}
  \subfloat[]{\label{fig:peps2b2}\includegraphics[width=0.3\textwidth]{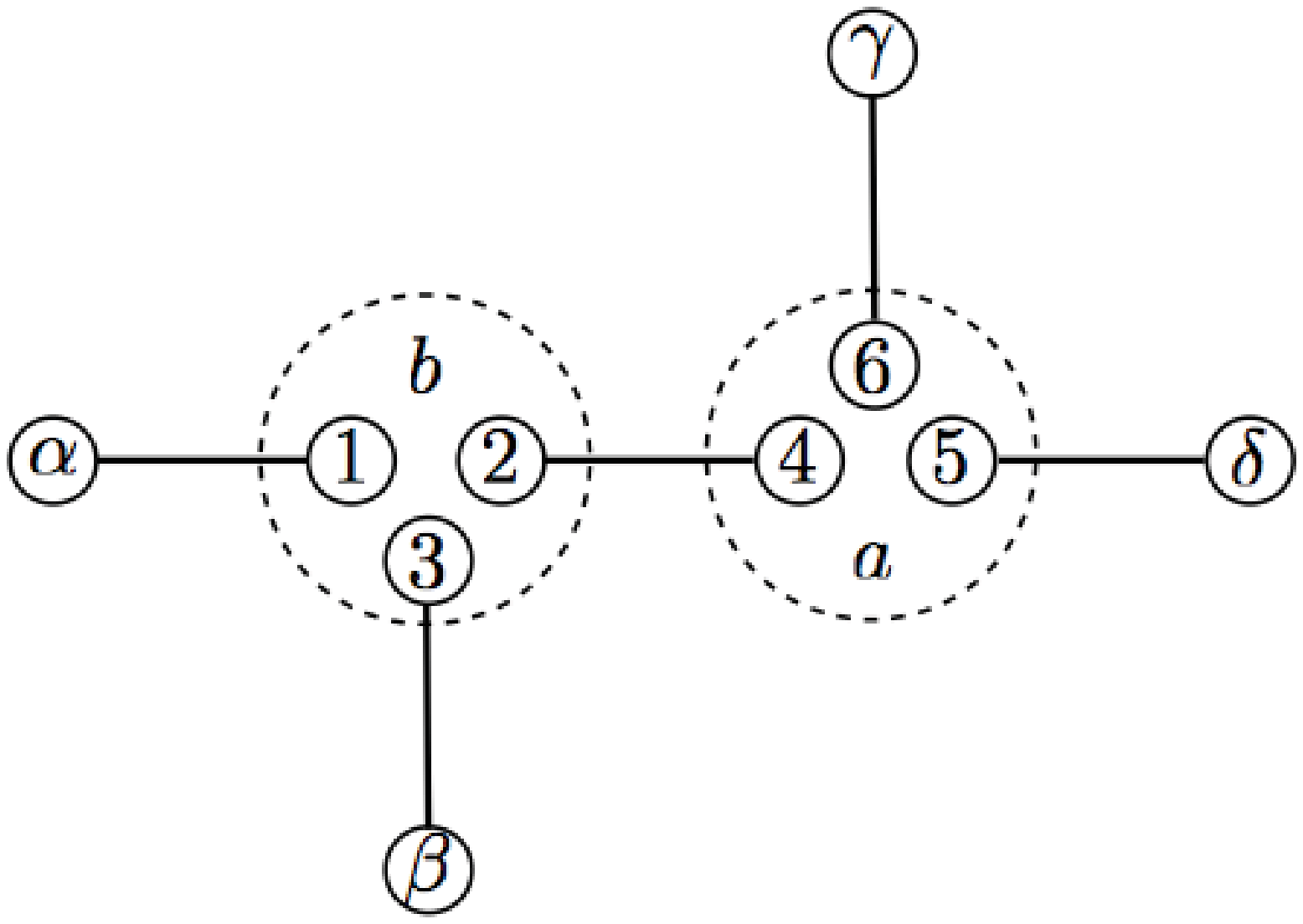}}
  \subfloat[]{\label{fig:peps2b3}\includegraphics[width=0.2\textwidth]{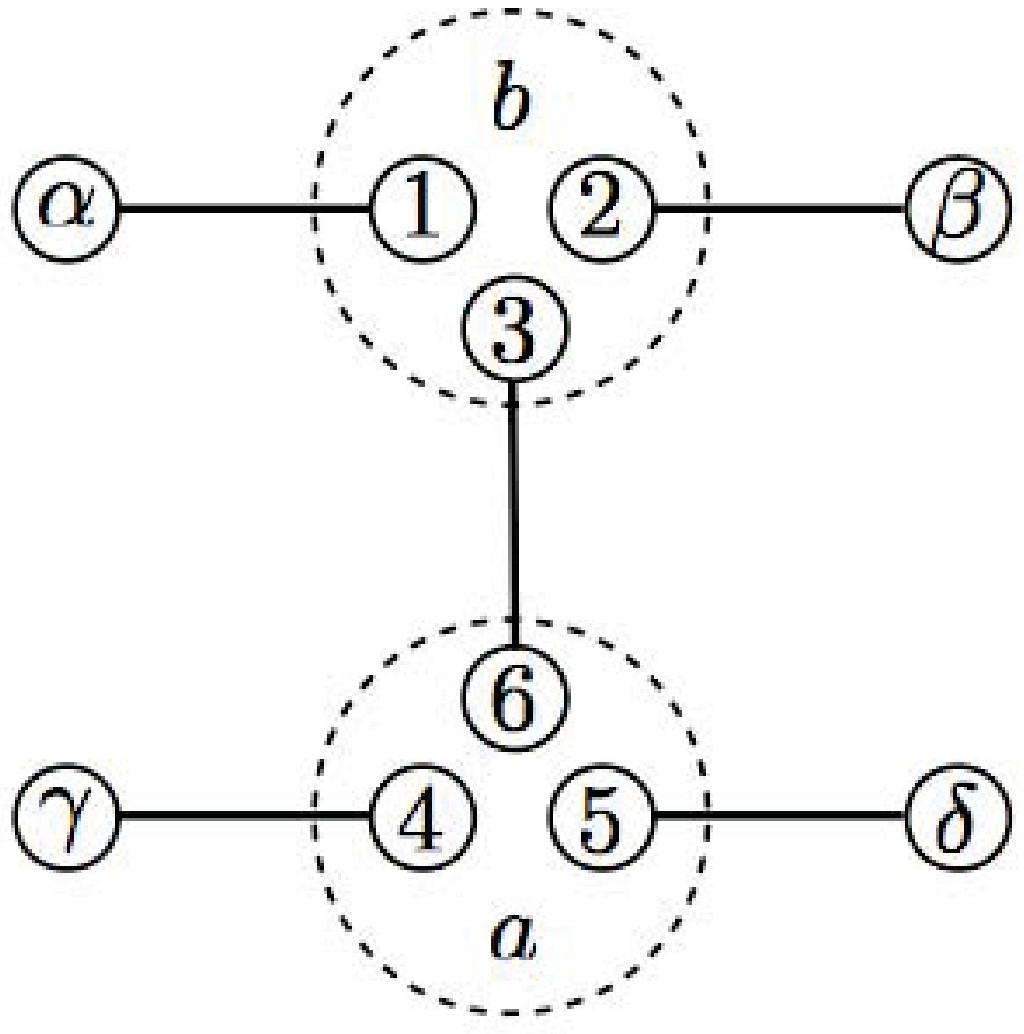}}
  \caption{Nearest-neighbor sites of the honeycomb lattice}
  \label{fig:peps2b}
\end{figure}

To construct $H_{triC}$, we start from the reduced density matrix of
$\ket{\Psi_{triC}}$ for any two neighboring spins. In general, these
two-particle reduced density matrices are dependent on the system
size, i.e., for different total number of spins $n$, these
two-particle reduced density matrices are different. However, the
range of these two-particle density matrices are independent of the
system size $n$. Here the range of a density matrix $\rho$ is the
space spanned by all the eigenvectors corresponding to nonzero
eigenvalues of $\rho$. Therefore, one can find a small system to
calculate the range of these two-particle density matrices. Because
the system has translational invariance, we only need to consider
three different kinds of neighbors, as shown in
Figs.~\ref{fig:peps2b1},~\ref{fig:peps2b2},~\ref{fig:peps2b3}.

In Fig.~\ref{fig:peps2b}, $a$ and $b$ in the circles refer to two
types of sites $a$ and $b$. Note that on the honeycomb lattice, the
sites of $a$ and $b$ are of different geometry, as one bond on site
$a$ goes up and one bond on site $b$ goes down. We call the
sublattice consisting of all the sites of $a$ type sublattice A and
the sublattice consisting of all the sites of $b$ type sublattice B.
Now denote the two-particle reduced density matrices corresponding
to the two spins in
Figs.~\ref{fig:peps2b1},~\ref{fig:peps2b2},~\ref{fig:peps2b3} by
$\rho_{ab}$, $\rho_{ba}$ and $\rho_{\stackrel{b}{a}}$ respectively,
and the corresponding ranges of $\rho_{ab}$, $\rho_{ba}$ and
$\rho_{\stackrel{b}{a}}$ by $S_{ab}$, $S_{ba}$ and
$S_{\stackrel{b}{a}}$ respectively. To compute $S_{ab}$, as shown in
Fig.~\ref{fig:peps2b1}, the virtual qubits 1 to 6 on those sites are
only connected to virtual qubits $\alpha$, $\beta$, $\gamma$,
$\delta$ elsewhere. By tracing out $\alpha$ to $\delta$ from the $5$
bonds $\ket{\psi_{cbond}}$, we get a $16$-dimensional space for
virtual qubits 1 to 6 spanned by
$|\pm\>_1|\pm\>_3|\psi_{singlet}\>_{24}|\pm\>_5|\pm\>_6$, where
$|\pm\>=(|0\>\pm |1\>)/\sqrt{2}$. This $16$-dimensional space is
then projected by $P_{triC}$s onto qubits $1,2,3$ and $4,5,6$
respectively to give $S_{ab}$, which is still a $16$-dimensional
space. Similarly, one can compute $S_{ba}$ and
$S_{\stackrel{b}{a}}$, which are all $16$-dimensional spaces.

Now we choose a two-body Hamiltonian involving only nearest-neighbor
interactions as the following
\begin{equation}
  H_{triC} = \sum_{a\in A} \left(h^p_{ab} + h^p_{ba} +
         h^p_{\stackrel{b}{a}}\right).
\label{H}
\end{equation}
Here the summation is over sites $a$ in sublattice $A$. And the
three terms $h^p_{ab}$, $h^p_{ba}$, $h^p_{\stackrel{b}{a}}$
correspond, respectively, to the projections onto the orthogonal
spaces of $S_{ab}$, $S_{ba}$ and $S_{\stackrel{b}{a}}$. Note the
overall Hilbert space of two spin-$5/2$ particles is $6\times
6=36$-dimensional, so $h^p_{ab}$, $h^p_{ba}$,
$h^p_{\stackrel{b}{a}}$ are all projections onto
$36-16=20$-dimensional spaces. Apparently, $\ket{\Psi_{triC}}$ is a
ground state of $H_{triC}$ as $h^p_{ab}\ket{\Psi_{triC}}=0$,
$h^p_{ba}\ket{\Psi_{triC}}=0$, and
$h^p_{\stackrel{b}{a}}\ket{\Psi_{triC}}=0$. Hence $H_{triC}$ is also
frustration-free.

To show that $|\Psi_{triC}\>$ is the unique ground state of $H_{triC}$, we need to
verify the condition that for any region $R$ of spins in $|\Psi_{triC}\>$,
the range $S_R$ of the reduced density matrix on $R$ satisfies
\begin{equation}
    S_R = \bigcap_{\< ab \>} S_{ab}\otimes I_{R\setminus ab},
\label{Unic}
\end{equation}
where the intersection is taken over all neighboring pairs $ab$
(i.e., including all the pairs as given in
Figs.~\ref{fig:peps2b1},~\ref{fig:peps2b2},~\ref{fig:peps2b3}), and
$I_{R\setminus ab}$ is the full Hilbert space of all spins in region
$R$ except $a$ and $b$~\cite{PVC+07}. For every possible
configuration containing three or four connected sites in
$|\Psi_{triC}\>$ the condition is confirmed by direct calculation.
For larger regions, this condition can be verified by
induction~\cite{CZG+09}, which is a directly application of Lemma 2
in~\cite{PVC+07}. Here we omit the technical parts of the results
in~\cite{PVC+07} and refer the readers there for more details.

We now show that $H_{triC}$ is also gapped, following the argument
in ~\cite{CZG+09}. Indeed, there exists a constant energy gap $\eta$
above the ground state. To estimate the value of $\eta$, we first
show that $\eta$ is greater than $\lambda$, the gap of another
Hamiltonian $K$ which also has $|\Psi_{triC}\>$ as its unique ground
state, but involves four-body interactions.

\begin{figure}[htb!]
\centering
\includegraphics[width=2.0in]{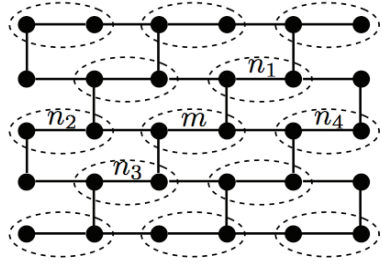}
\caption{(This figure is redrawn from FIG. 3 in ~\cite{CZG+09}.)
Regrouping of lattice sites in tri-Cluster state into disjoint
blocks, each containing two sites.} \label{fig:HX}
\end{figure}

Consider a Hamiltonian $K$ for a re-labeled version of
$|\Psi_{triC}\>$, in which particles are regrouped into disjoint
blocks with each containing two nearest neighbors, as shown in
Fig.~\ref{fig:HX}.
Note as discussed above, $K$ also has $|\Psi_{triC}\>$ as its unique
ground state, where the corresponding condition similar to
Eq.~\eqref{Unic} can be verified by induction, based on Lemma 2
in~\cite{PVC+07}.

Now let $K = \sum_{mn} k_{mn}$, where $m,n$ denote two connected
blocks, each containing two particles $m^{[l]},m^{[r]}$ and
$n^{[l]},n^{[r]}$ respectively. Assuming $m^{[r]}$ and $n^{[l]}$
are connected, $k_{mn}$ is then the projection onto the
orthogonal space of the four-particle reduced density matrix of particles
$m^{[l]},m^{[r]},n^{[l]},n^{[r]}$. We then have
\begin{equation}
\label{eq:K}
H_{triC} = \sum_{ab} h^p_{ab} \ge \frac{1}{4}
\sum_{mn} \left(h^p_{m^{[l]}m^{[r]}} + h^p_{m^{[r]} n^{[l]}} +
h^p_{n^{[l]}n^{[r]}}\right) \ge \frac{1}{4} \sum_{mn} \mu k_{mn} =
\frac{1}{4}\mu K.
\end{equation}

Note that both $\left(h^p_{m^{[l]}m^{[r]}} + h^p_{m^{[r]} n^{[l]}}
+ h^p_{n^{[l]}n^{[r]}}\right)$ and $k_{mn}$ are non-negative
operators with the same null space, so the second inequality holds
in Eq.~\eqref{eq:K} for
some positive number $\mu$.  Without loss of generality,
one can assume that the gaps of the projectors
$h^p_{ab}$ and $k_{mn}$ are both $1$, then  direct calculation gives
$\mu=\frac{1}{2}$.  One then has
$\eta \ge \frac{1}{4}\mu \lambda = \frac{1}{8} \lambda$.

We then bound the gap $\lambda$. To do this, one can show that $K^2
\ge c K$ for some positive constant $c$. This is given by direct
calculation of $K^2$ and compare it with $K$, which finally gives
$c=1/3$. Therefore, finally one finds a lower bound on the gap
$\eta$ of $H_{triC}$ with $\eta \ge \frac{1}{8} \lambda \ge
\frac{1}{24}$.

\section{Resource states in spin-$3/2$ systems}
\label{sec:spinthreehalf}

After showing that realistic resource states for MBQC could be found
in spin-$5/2$ systems, in this section we will see that we could go
further, i.e., quantum states of spin-$3/2$ systems are also
possible to serve as realistic resource states.

\subsection{The 2D AKLT state}
\label{sec:AKLT2D}

In Sec.~\ref{sec:1D}, we have shown that the 1D AKLT state can be
reduced to a 1D cluster state, thus it can be a resource state of
one-qubit MBQC. One would then naturally ask whether any 2D AKLT
state can be a resource state for MBQC. Recently, Wei {\it et al.}
and Miyake showed independently that the 2D AKLT state on the
honeycomb lattice, which is a system composed of spin-$3/2$
particles, is a universal resource state for
MBQC~\cite{Miy10,WAR11}.

We briefly discuss the approach by Wei {\it et al.}. Before starting
the discussion, let us explain a useful concept, namely quantum
encoding (see Chap 10 of \cite{NC00} for more details). Remember
that if a quantum system is two-dimensional, we may call it a qubit.
Naturally, if a quantum system is composed of more than one qubits,
the dimension of its state space will be higher. However, it is
possible that sometimes only a two-dimensional subspace of the state
space of this many-qubit system is involved in our consideration. In
this situation, for convenience we often regard the entire system as
one qubit, and we say that the many-qubit system is encoded into one
encoded qubit or one logical qubit. Similarly, the corresponding
Pauli matrices of the encoded qubit are called logical Pauli
matrices. Furthermore, in a complicated system we might encode
different subsystems in different ways, and besides, the procedure
of encoding might also be iterated. As a rule, if a quantum state
after encoding is a cluter state, we will call this state an encoded
cluster state.

It is discussed in Sec.~\ref{sec:1D} that the 1D AKLT state of
spin-$1$'s has a VBS representation. Similarly, 2D AKLT states can
also be considered in this manner~\cite{KX10}. Indeed, 2D AKLT
states can be defined on any kind of 2D lattices, or even an
arbitrary graph~\cite{KX10}. For instance, if one chooses the
projection in Fig.~\ref{fig:cluvbs2} as the one onto the symmetric
subspace of the four virtual qubits, the state will be a 2D AKLT
state on a square lattice, and the corresponding particles are
spin-$2$; if the projection in Fig.~\ref{fig:cluvbs3} is also onto
the symmetric subspace, but of the three virtual qubits,  the state
will be a 2D AKLT state on the honeycomb lattice, and the
corresponding particles are spin-$3/2$. Note that though similar,
the two symmetric subspace is different in dimension, i.e., the
former on the square lattice is of dimension $5$ (spin-$2$), and the
latter on the honeycomb lattice is of dimension $4$ (spin-$3/2$).

Let us go back to the work of Wei {\it et al.} Based on the
observation discussed in Sec.~\ref{sec:resource}, their basic idea
is to prove that the 2D AKLT state on the honeycomb lattice can be
converted to an encoded cluster state on a planar lattice by
adaptive local measurements. More concretely, firstly they found a
generalized measurement and perform it on every spin-$3/2$ particle.
Secondly, they showed that there exists an encoding scheme
determined by the measurement outcomes such that the AKLT state on
the honeycomb can be regarded an encoded cluster state on a random
planar graph (i.e., a graph that can be embedded in the plane).
Finally, through numerical simulation and percolation theory they
demonstrated that a typical resulting graph state is universal for
MBQC.

\begin{figure}
  \begin{center}
 \includegraphics[width=3.5in]{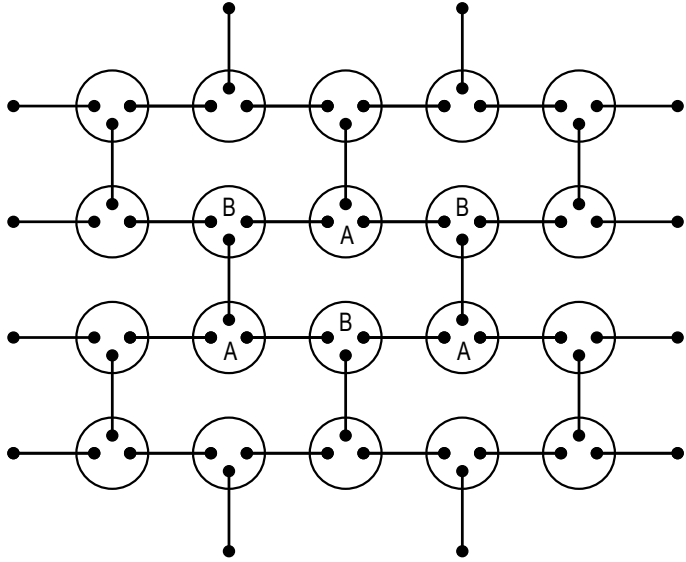}
    \end{center}
  \caption{\label{fig:2daklt} 2D AKLT state on the honeycomb lattice. Note that all the sites are divided into part A and part B.}
\end{figure}

As mentioned above and indicated by Fig.~\ref{fig:2daklt}, in every
site (shown as a circle) of the AKLT state on the honeycomb lattice $\mathcal{L}$
locates one spin-$3/2$ particle. Being a VBS state, every spin-$3/2$
of $\mathcal{L}$ can be viewed as the four-dimensional symmetric
subspace of three virtual qubits. The corresponding projection
onto this subspace for site $v$ is
\begin{equation}\label{eq:2Dproj}
P_{S,v}=|\tilde{0}\rangle\langle000|+|\tilde{1}\rangle\langle111|+|\tilde{2}\rangle\langle
W|+|\tilde{3}\rangle\langle\bar{W}|,
\end{equation}
where
$|W\rangle=\frac{1}{\sqrt{3}}(|001\rangle+|010\rangle+|100\rangle)$
and
$|\bar{W}\rangle=\frac{1}{\sqrt{3}}(|110\rangle+|101\rangle+|011\rangle)$.
Thus, the AKLT state on $\mathcal{L}$ can be expressed as
\begin{equation}
|\Phi_{AKLT}\rangle\equiv\bigotimes_{v\in
V(\mathcal{L})}P_{S,v}\bigotimes_{e\in
E(\mathcal{L})}|\psi_{singlet}\rangle_e,
\end{equation}
where $V(\mathcal{L})$ and $E(\mathcal{L})$ are the sets of vertices
and edges respectively, and $|\psi_{singlet}\rangle_e$ is a singlet
state defined in Eq.(\ref{eq:singlet}) with its qubits at the two
sites connected by $e$.

For any site $v$, consider the following three projections,
\begin{eqnarray}
\label{eq:proj}
F_{v,z}&=&\sqrt{\frac{2}{3}}(|000\rangle\langle000|+|111\rangle\langle111|),\nonumber\\
F_{v,x}&=&\sqrt{\frac{2}{3}}(|+++\rangle\langle+++|+|---\rangle\langle---|),\nonumber\\
F_{v,y}&=&\sqrt{\frac{2}{3}}(|+i,+i,+i\rangle\langle
+i,+i,+i|+|-i,-i,-i\rangle\langle-i,-i,-i|),
\end{eqnarray}
where $|\pm\rangle=(|0\rangle\pm|1\rangle)/\sqrt{2}$, and $|\pm
i\rangle=(|0\rangle\pm i|1\rangle)/\sqrt{2}$. It can be checked that
\begin{equation}\label{eq:AKLTpro}
\sum_{k\in\{x,y,x\}}F^\dag_{v,k}F_{v,k}=P_{S,v}.
\end{equation}
Namely, these projections form a generalized measurement on each
site of the lattice. The key part of the work by Wei el al. is that
they proved if one performs this generalized measurement on every
site of the lattice, the resulting state will be an encoded cluster
state on some random planar graph, up to local unitary operations. Suppose the set of
measurement outcomes are $\mathcal{A}=\{a_v, v\in V(\mathcal{L})\}$,
then the resulting state will be a function of $\mathcal{A}$, which
is denoted by $|\Psi(\mathcal{A})\rangle$. According to
Eq.~\eqref{eq:AKLTpro}, we have
\begin{equation}
|\Psi(\mathcal{A})\rangle=\bigotimes_{v\in
V(\mathcal{L})}F_{v,a_v}|\Phi_{AKLT}\rangle=\bigotimes_{v\in
V(\mathcal{L})}F_{v,a_v}\bigotimes_{e\in
E(\mathcal{L})}|\psi_{singlet}\rangle_e.
\end{equation}

\begin{figure}
  \begin{center}
 \includegraphics[width=4in]{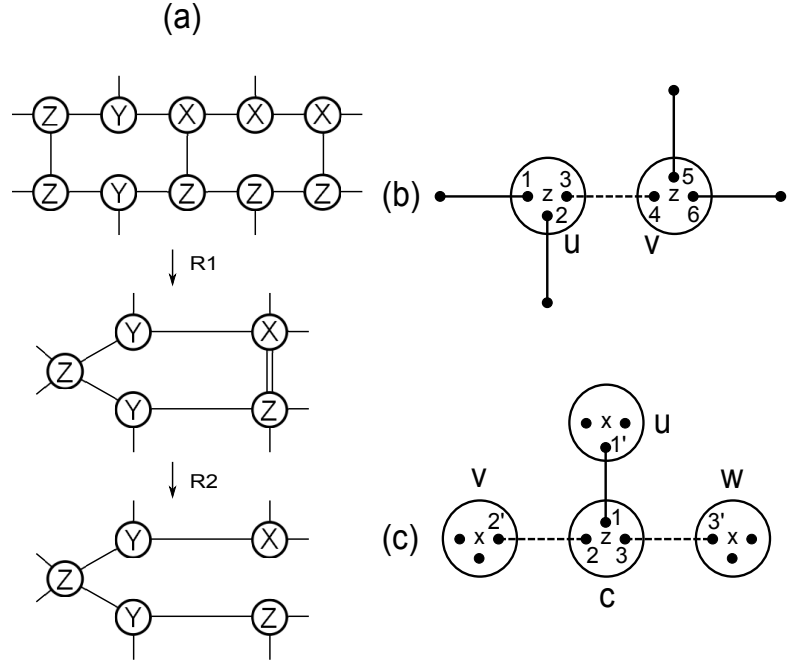}
    \end{center}
  \caption{\label{fig:2dencoding} [(b) and (c) of this figure are redrawn from (c) and (d) of FIG. 2 in ~\cite{WAR11}.]
  (a) Rules R1 and R2. (b) A domain containing two sites. (c) A domain is connected to three other domains, and every domain is supported by one site.}
\end{figure}

Wei {\it et al.} gave a constructive proof to show that
$|\Psi(\mathcal{A})\rangle$ is an encoded cluster state on some
random planar graph, up to local unitary operations. Firstly, as a
planar graph, the lattice of spin-$3/2$ particles can be recast by
two rules illustrated in Fig.~\ref{fig:2dencoding}(a). These two
rules are: (R1) (Edge contraction): Contract all edges $e\in
E(\mathcal{L})$ that connect sites with the same measurement
outcomes. (R2) (Mod-2 edge deletion): In the resultant multigraph,
delete all edges of even multiplicity and convert all edges of odd
multiplicity into edges with multiplicity 1. If several sites of
$\mathcal{L}$ is contracted into one single site by R1, we call
these sites a domain. Every domain will be an encoded qubit in the
target encoded cluster state.

Let us see why a domain can encode a qubit. If the domain contains
only one site, the form of the measurement suggests that the state
of every domain will be in a two-dimensional subspace, which naturally means a
qubit. The case that one domain contains more than one
site can be illustrated by an example in
Fig.~\ref{fig:2dencoding}(b), where one domain is composed by two
adjacent sites $u$ and $v$ with the same measurement outcome $a_z$.
Considering the form of $F_{v,z}$ in Eq.~\eqref{eq:proj}, it can be
known that the operators $Z_1Z_2$, $Z_2Z_3$, $Z_4Z_5$, and $Z_5Z_6$
are stabilizer elements of $|\Psi(\mathcal{A})\rangle$. Here we use
the framework of the stabilizer theory proposed by D. Gottsman
\cite{Gottes97}. For a quantum state $|\alpha\rangle$ and a operator
$A$, if $A|\alpha\rangle=|\alpha\rangle$, we say $|\alpha\rangle$ is
stabilized by $A$, and $A$ is a stabilizer element of
$|\alpha\rangle$, where $A$ is usually chosen from Pauli operators.
Actually, in Sec.~\ref{sec:MBQC} we have met some
examples, and Eq.(\ref{eq:eig1}) in Sec.~\ref{sec:MBQC} is one of them, which shows
some stabilizer elements of the cluster state $\ket{\Psi_C}$. Interestingly, it can be
proved that the corresponding cluster state is the only state that
is stabilized by all the operators of the form given by
Eq.(\ref{eq:eig1})~\cite{NC00}, which demonstrates the purpose of
introducing the stabilizer theory, i.e., it is possible to
characterize a quantum state by its stabilizer elements. Similarly,
noting that $-Z_3Z_4$ commutes with $F_{u,z}\otimes F_{v,z}$, and
that $|\psi_{singlet}\rangle_{34}$ is stabilized by $-Z_3Z_4$, one
can obtain that $-Z_3Z_4$ is also a stabilizer element of
$|\Psi(\mathcal{A})\rangle$. In this way, the state of this domain
could be expressed as
\begin{equation}
\alpha|(000)_u(111)_v\rangle+\beta|(111)_u(000)_v\rangle,
\end{equation}
which therefore might serve as an encoded qubit if one chooses the
basis states as $|\bar{0}\rangle=|(000)_u(111)_v\rangle$ and
$|\bar{1}\rangle=|(111)_u(000)_v\rangle$.

Using the similar idea, we could consider a general domain by the
stabilizer theory. Suppose there are $|\mathcal{C}|$ sites in this
domain, and at present we still suppose the corresponding
measurement outcome is $a_z$. Then $|\Psi(\mathcal{A})\rangle$ is
stabilized by $\{\lambda_i\lambda_jZ_iZ_j,
i,j=1,2,...,3|\mathcal{C}|\}$. Here $\lambda_i=1$ if and only if
$i\in v\in A$, and $\lambda_i=-1$ otherwise (note that the 2D AKLT
state on the honeycomb could be divided into part A and part B as in
Fig.~\ref{fig:2daklt}). The reason that $\lambda_i$'s are introduced
is the form of the stabilizer elements of $|\psi_{singlet}\rangle$,
which has a negative sign for $Z_iZ_j$. As an encoded qubit, one can
introduce the logical $X$ operation as
$\bar{X}=\otimes_{j=1}^{3|\mathcal{C}|}X_j$, and the logical $Z$
operation as $\bar{Z}=\lambda_iZ_i$. Similarly, for domains with
different measurement outcomes, one can also find proper stabilizer
elements and logical operations. For example, if the outcome is
$a_y$, one could choose the stabilizer generators as
$\{\lambda_i\lambda_jY_iY_j, i,j=1,2,...,3|\mathcal{C}|\}$,
$\bar{X}$ as $\otimes_{j=1}^{3|\mathcal{C}|}Z_j$ and $\bar{Z}$ as
$\lambda_iY_i$; if the outcome is $a_x$, they can be chosen as
$\{\lambda_i\lambda_jX_iX_j, i,j=1,2,...,3|\mathcal{C}|\}$,
$\otimes_{j=1}^{3|\mathcal{C}|}Z_j$, and $\lambda_iX_i$
respectively.

When one domain occupies more than one sites, the domain and thus
the encoding can be simplified. By measurements on those redundant
sites, every logical qubit can be supported by one site only. For
instance, if one measures the site $v$ in the logical qubit of
Fig.~\ref{fig:2dencoding}(b) in the basis
$\{|(000)_v\rangle\pm|(111)_v\rangle\}$, the resulting state will
become $\alpha|(000)_u\rangle+\beta|(111)_u\rangle$.

After determining the encoding scheme, we have to show why the
encoded state obtained is indeed a cluster state, which can be
demonstrated by the example in Fig.~\ref{fig:2dencoding}(c), where
the state is composed by four domains, and each domain contains one
site only. Consider the operator
$\mathcal{O}=-X_1X_{1'}X_2X_{2'}X_3X_{3'}$, and we would like to
prove that it is in stabilizer of $|\Psi(\mathcal{A})\rangle$. In
fact, according to the form of $|\psi\rangle_e$, it is easy to know
that for any $i\in\{1,2,3\}$ $-X_iX_{i'}$ is in stabilizer of
$\otimes_{e\in E(\mathcal{L})}|\psi\rangle_e$. Besides, for any
$i\in\{1,2,3\}$, $-X_iX_{i'}$ commutes with $F_{k,a_x}$ for any
$k\in\{u,v,w\}$. Therefore, in order to prove
$|\Psi(\mathcal{A})\rangle$ is stabilized by
$\mathcal{O}=-X_1X_{1'}X_2X_{2'}X_3X_{3'}$, one only need to show
that this operator commutes with $F_{c,a_z}$, which can be verified
easily.

On the other hand, as logical single qubits, we have stipulated
logical Pauli operators for these four domains, which include
$\bar{X_c}=Z_1Z_2Z_3$, $\bar{Z_u}=\pm X_{1'}$, $\bar{Z_v}=\pm
X_{2'}$, and $\bar{Z_w}=\pm X_{3'}$. As a result, it holds that
$\mathcal{O}=\pm\bar{X_c}\bar{Z_u}\bar{Z_v}\bar{Z_w}$. Up to some
signs, this is exactly the key character of cluster states, as we have
introduced in Sec.3 (see Eq.\ref{eq:eig1}).

It should be pointed out that when applying Rule (R1), multiple
edges might be produced. As an instance, if there exists a
$m$-multiplicity ($m>1$) edge between domains $w$ and $v$, one will
find the resulting state is stabilized by $\bar{X_w}\bar{Z_v}^m$,
which is actually $\bar{X_w}\bar{Z_v}^{(m\mod 2)}$ because of the fact
$\bar{Z_v}^2=I$. Therefore, one can find out that the introduction
of Rule (R2) will simplify the graph while the corresponding
resulting state is kept unchanged.

To prove that 2D AKLT state on the honeycomb is a universal resource
state for MBQC after getting the encoded cluster states, one still
needs to make sure that the connectivity properties of the resulting
cluster state is good enough. In Ref.~\cite{WAR11}, Wei {\it et al.}
provided convincing numerical evidences to show that this is indeed
the case.

It should be pointed out that recently Darmawan {\it et al.} showed
that most states in the disordered phase of a large family of
Hamiltonians characterized by short-range-correlated VBS states
could be reduced to cluster states that are universal for MBQC, and
the 2D AKLT state on the honeycomb is just an example~\cite{DBB11}.

\subsection{The Quasi-AKLT Chains and Quantum Magnets}

%%This subsection reviews the results in~\cite{CMD+10} and~\cite{WRK11}.
In the subsection above, a nice result that by local operations 2D
AKLT states on the honeycomb can be converted to 2D graph states is
elaborated. As a realistic resource state for MBQC, however, one
link is still missing for this 2D AKLT. Though widely believed, it
remains unknown whether the 2D AKLT Hamiltonian on the honeycomb
lattice is gapped~\cite{AKLT88}. That is, Condition $4$ given in
Sec.~\ref{sec:resource} might not be satisfied. In this subsection,
we discuss another work by Cai {\it et al.}. Based on a new defined
system, namely 1D AKLT quasichains of mixed spin-$3/2$'s and
spin-$1/2$'s, they constructed an ideal resource state in a
spin-$3/2$ system for MBQC, which successfully satisfies all of
Conditions $1-5$ given in Sec.~\ref{sec:resource}~\cite{CMD+10}.
This work was later reinterpreted by Wei {\it et al.} based on the
observation in $4$ given in Sec.~\ref{sec:resource} for universality
and the technique of generalized measurement discussed in the above
subsection~\cite{WRK11}. For convenience and consistency, we discuss
the approach of~\cite{WRK11} to introduce the basic idea
of~\cite{CMD+10}.

\begin{figure}
  \begin{center}
 \includegraphics[width=11cm]{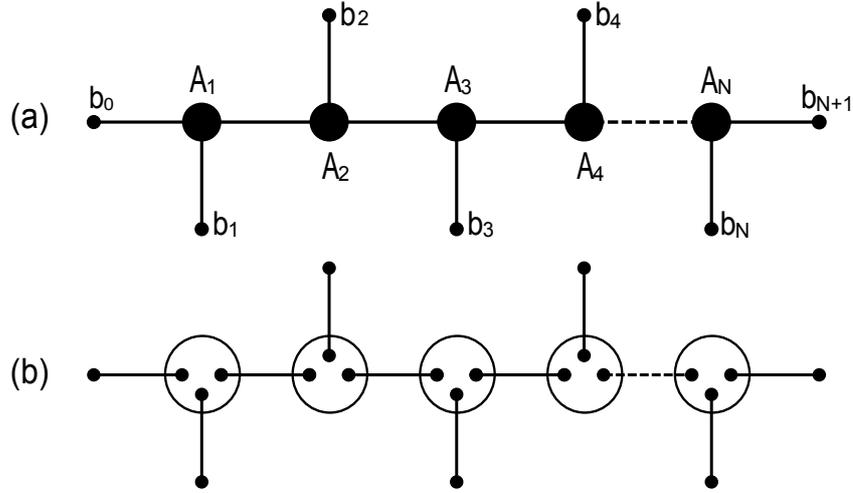}
    \end{center}
  \caption{\label{fig:fig1D} (This figure is redrawn from FIG. 1 in ~\cite{WRK11}.) (a) 1D AKLT quasichain. (b) The ground state of 1D AKLT quasichain is a VBS state.}
\end{figure}

An AKLT quasichain was defined by Cai {\it et al.} as in
Fig.~\ref{fig:fig1D}(a). A little different from the original AKLT
chain, an AKLT quasichain is composed mainly by some spin-$3/2$
particles forming a backbone, and every spin-$3/2$ is attached by
one or two spin-$1/2$'s. As discussed before, every spin-$3/2$ can
be explained as three virtual spin-$1/2$'s followed by the
projection onto the symmetric subspace, which is shown by
Fig.~\ref{fig:fig1D}(b), and the corresponding state can be
expressed as
\begin{equation}
|\Psi_{AKLT}\rangle\sim \bigotimes_AP_{S,A}\bigotimes_{e\in
edge}|\psi_{singlet}\rangle_e,
\end{equation}
where $|\psi_{singlet}\rangle_e=(|01\rangle-|10\rangle)/\sqrt{2}$ is
a singlet state on $e$, and $P_{S,A}$ is defined as
Eq.~\ref{eq:2Dproj}. The Hamiltonian of the quasichain is
\begin{equation}
H=J\left(\sum_{i=1}^{N-1}P_{A_i,A_{i+1}}^{S=3}+\sum_{i=1}^{N}P_{A_i,b_i}^{S=2}+P_{A_1,b_0}^{S=2}+P_{A_N,b_{N+1}}^{S=2}\right),
\end{equation}
where $P_{a,b}^{S=k}$ is the operator projecting the total spin of
$a$ and $b$ onto the spin-$k$ subspace. It can be shown that a AKLT
quasichain has a non-degenerate ground state with a finite energy
gap~\cite{CMD+10}.

To be universal for MBQC, 2D model is needed. Based on 1D AKLT
quasichains, Cai {\it et al.} constructed an interesting 2D gapped
system and proved that its unique ground state is universal for
MBQC. This 2D model could be illustrated by Fig.~\ref{fig:fig2D}.
The key technique in the construction is to combine two spin-$1/2$'s
from two neighboring 1D AKLT quasichains into one spin-$3/2$. The
combination could be expressed as a unitary transformation $U$,
\begin{equation}\label{eq:mapping}
U=\sum_{m_1,m_2=\pm1/2}\Big|\frac{3}{2},m_1+2m_2\Big\rangle_B\Big\langle\frac{1}{2},m_1\Big|_{b_1}\Big\langle\frac{1}{2},m_2\Big|_{b_2},
\end{equation}
which is actually a relabeling of the basis states. Importantly,
note that under this unitary transformation both unitary operations
and measurements on spin-$1/2$'s $b_1$ and $b_2$ are essentially
corresponding operations on the spin-$3/2$ particle $B$. In this
way, the main part of this 2D model is spin-$3/2$ particles, and the
only exception is the spin-$1/2$'s on the boundary, which actually
can also be avoided by considering periodic boundary conditions. The
Hamiltonian of this 2D model can be expressed as
\begin{equation}
\label{eq:H2D}
H'=\mathcal{U}\left(\sum_kH^{k}\right)\mathcal{U}^{\dag},
\end{equation}
where $H^{k}$ is the Hamiltonian of the $k$-th quasichain, and
$\mathcal{U}$ is the tensor product of all necessary merging
operations given by Eq.(\ref{eq:mapping}). Cai {\it et al.} showed
that $H'$ is also gapped~\cite{CMD+10}.

By the same generalized measurement given in Eq.\eqref{eq:proj} and similar encoding approach as
introduced in Sec.~\ref{sec:AKLT2D}, the ground state of the AKLT
quasichain can be converted to a 1D cluster state by local
operations. More precisely, one has to perform the measurement
defined by Eq.\eqref{eq:proj} on every spin-$3/2$ particle, then the resulting
state turns out to be an encoded cluster state, where as discussed before the
encoding scheme is determined by the measurement outcomes.
Meanwhile, a subtle point here is that some
spin-$1/2$'s exist. We have mentioned that these spin-$1/2$'s are
critical in increasing the dimension of the resulting cluster state
from 1 to 2. For this 1D case, since the procedure of encoding is
almost a repeat of the one in Sec.~\ref{sec:AKLT2D}, we provide the
encoding details directly.

\begin{figure}
\begin{center}
  \includegraphics[width=3.5in]{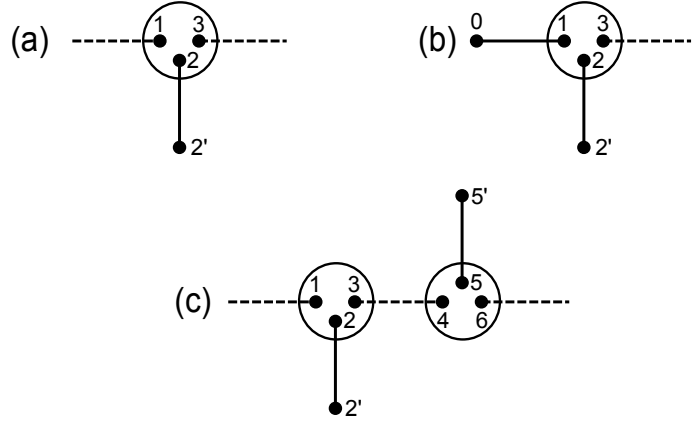}
\end{center}
  \caption{\label{fig:figEncoding} (This figure is redrawn from FIG. 3 in ~\cite{WRK11}.) Encoding structure of a logical qubit. (a) One spin-$3/2$ and one spin-$1/2$ encode into a logical qubit.
  (b) One spin-$3/2$ and two spin-$1/2$'s encode into a logical qubit. (c) Two consecutive spin-$3/2$'s and two spin-$1/2$'s attached to them encode
  into a logical qubit.}
\end{figure}

As shown in Fig.~\ref{fig:figEncoding}(a), suppose the measurement
outcome on site $u$ is $a_z$, then the logical basis states are
$|\bar{0}\rangle=|(000)_u1_{2'}\rangle$ and
$|\bar{1}\rangle=|(111)_u0_{2'}\rangle$, and the logical Pauli
operations could be chosen as $\bar{Z}=Z_1$ and $\bar{X}=X_1\otimes
X_2\otimes X_3\otimes X_{2'}$. Similarly, if the measurement out is
$a_x$, then these states and operations are
$|\bar{0}\rangle=|(+++)_u(-)_{2'}\rangle$,
$|\bar{1}\rangle=|(---)_u(+)_{2'}\rangle$, $\bar{Z}=X_1$ and
$\bar{X}=Z_1\otimes Z_2\otimes Z_3\otimes Z_{2'}$. The case of
measurement outcome being $a_y$ is also similar. Note that it is
possible that some spin-$3/2$ is connected to 2 spin-$1/2$'s, as in
Fig.~\ref{fig:figEncoding}(b). In this case, the encoding is the
same except that the state of spin-$1/2$ needs to be repeated once.
For instance, when the measurement outcome is $a_z$, the logical
$|0\rangle$ becomes $|\bar{0}\rangle=|(000)_u1_01_{2'}\rangle$. The
third possibility of encoding a logical qubit is that several
consecutive spin-$3/2$ particles have the same measurement outcome, as
illustrated in Fig.~\ref{fig:figEncoding}(c), where the two logical
basis states could be $|(000)_u1_{2'}(111)_v0_{5'}\rangle$ and
$|(111)_u0_{2'}(000)_v1_{5'}\rangle$. Under the encoding, the ground
state of every 1D AKLT quasichain can be proved to be 1D cluster
state, which can be verified by the stabilizer theory as in Sec.~\ref{sec:AKLT2D}.
Furthermore, since one could simplify every
logical qubit by measuring the unwanted spin-$3/2$'s and
spin-$1/2$'s, it is reasonable to assume that every logical qubit
occupies only one spin-$3/2$, which has also be similarly explained in Sec.~\ref{sec:AKLT2D}.
For example, the basis state $|\bar{0}\rangle$ in
Fig.~\ref{fig:figEncoding}(a) could be simplified from
$|(000)_u1_{2'}\rangle$ to $|(000)_u\rangle$ by measuring qubit $2'$
in basis $(|0\rangle\pm|1\rangle)/\sqrt{2}$.

After knowing how to obtain logical 1D cluster states, the main
challenge remaining is to produce a 2D cluster state by somehow
connecting them together. In order to overcome this, Cai {\it et
al.} proposed the following interesting solution, as illustrated by
Fig.~\ref{fig:fig2D}.

\begin{figure}
\begin{center}
   \includegraphics[width=3.5in]{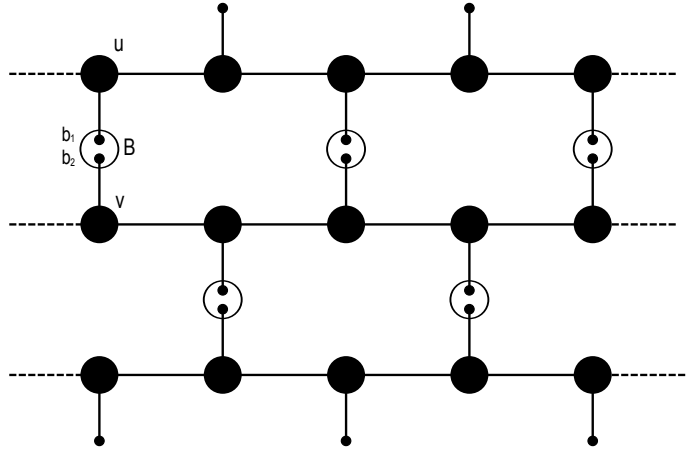}
\end{center}
  \caption{\label{fig:fig2D} 2D model based on 1D AKLT quasichains. Spin-$1/2$ pairs in the circles are mapped into spin-$3/2$'s. }
\end{figure}

To produce universal resource states for MBQC, for arbitrary two
neighboring logical qubits $u$ and $v$ on different quasichains, one
needs the capability to perform two possible operations between
them: One is a logical controlled-$Z$ gate defined in
Eq.\eqref{eq:cZ}, and the other is to keep the state of every
logical qubit unchanged. Cai {\it et al.} proved that both of them
could be realized by operations on qubits $b_1$ and $b_2$ only, up
to local operations on logical qubits $u$ and $v$. Recall that
unitary operators and measurements on qubits $b_1$ and $b_2$ are
essentially corresponding operations on the spin-$3/2$ particle $B$.

Without loss of generality, we assume that when producing 1D cluster
states, both of the measurement outcomes on sites $u$ and $v$ are
$a_z$. To keep the two logical qubits unchanged, one first applies
two $X$ operations on qubits $b_1$ and $b_2$, then measures them in
basis $(|0\rangle\pm|1\rangle)/\sqrt{2}$ separately. If the new
measurement outcomes are $m_1$ and $m_2$, it can be checked that the
effective operation is to perform $Z_u^{m_1}$ and $Z_v^{m_2}$ on
logical qubits $u$ and $v$ respectively. Actually, this is to
essentially simplify the encoding of logical qubits $u$ and $v$ by
removing the unwanted spin-$1/2$'s, and the resulting state might
need the correction of local unitary operations.

Let us turn to the case of controlled-$Z$ operation. Once more, in
this case two X operations are needed on qubits $b_1$ and $b_2$
firstly, then a controlled-$Z$ gate is applied on these two qubits.
Up to local operations, a logical controlled-$Z$ gate is
implemented.

In short, we see that 1D AKLT quasichains can be connected together
and then be converted to a logical 2D cluster state, which is
universal for MBQC. By showing that the operations needed for
related spin-$1/2$'s correspond to operations on the resulting
spin-$3/2$'s state through the mapping in Eq.(~\ref{eq:mapping}), we
see that that the ground state of the 2D model defined by Cai {\it
et al.} can be a realistic resource state for MBQC.

\section{Universal MBQC in spin-$1/2$ systems}
\label{sec:spinonehalf}

In the above sections, we have seen successful examples showing the
existence of spin-$5/2$ and spin-$3/2$ systems that are ideal
resource states for MBQC. As the most practical quantum systems, it
is natural to ask whether it is possible to find ideal resource
states for MBQC from qubit systems. Unfortunately, it has been
pointed out by Chen {\it et al.} that this is not the
case~\cite{CCD+10}.

The basic idea of Chen {\it et al.} is to show that for any two-body
frustration-free qubit Hamiltonian, there always exists a ground
state which is a product of single- or two-qubit states. That is to
say, any entangled state cannot be the non-degenerate ground state
of such a Hamiltonian, which rules out the possibility of qubit
systems being ideal resource states for MBQC.

Consider an $n$-qubit state $|\Psi\rangle$ which is genuinely entangled.
That is, $|\Psi\rangle$ cannot be a product state with respect to any bipartition of these $n$ qubits.
We denote the reduced density matrix of qubits $i$ and $j$ by
$\rho_{ij}$. Consider the following Hamiltonian $H_{\Psi}$
\begin{equation}
\label{eq:HFF}
H_{\Psi}=\sum_{ij}\Pi_{ij},
\end{equation}
where $\Pi_{ij}$ is the projection onto the kernel of $\rho_{ij}$.

It can be observed that $H_{\Psi}$ is a two-body frustration-free
qubit Hamiltonian, and $|\Psi\rangle$ is in $S(\Psi)$, the ground
space of $H_{\Psi}$. Actually, one can find that $S(\Psi)$ is the
smallest among all the ground-state spaces of two-body
frustration-free qubit Hamiltonians that contain $|\Psi\rangle$. For
a general two-body frustration-free qubit Hamiltonian
$H=\sum_{k}H_k$, $H_k$'s might not be projections. However, we could
replace them with $\Pi'_{k}$s and the ground-state space does not
change, where $\Pi_{k}$ is the projection onto the orthogonal
subspace of the ground-state space of $H_k$. In this way, we only
need to consider Hamiltonians composed of projections, which are of
the form of Eq.\eqref{eq:HFF}.

Next, let us note an important point that helps with the proof.
Suppose there exist $2\times2$ nonsingular linear operators
$L_1,...,L_n$, and $|\Psi\rangle=L|\Phi\rangle$, where
$L=L_1\otimes\cdots\otimes L_n$. In the language of quantum
information theory, two states $|\Phi\rangle$ and $|\Psi\rangle$ can
be transformed to each other via $L$ are called equivalent under
stochastic local operation and classical communication
(SLOCC)\cite{BPR+00}. Note that $|\Psi\rangle$ is a ground state of
$H$ if and only if $|\Phi\rangle$ is a ground state of
\begin{equation}
H'=\sum_{ij}(L_i\otimes L_j)^{\dag}\Pi_{ij}(L_i\otimes L_j),
\end{equation}
and that if one of $|\Psi\rangle$ and $|\Phi\rangle$ is product
state, the other one also is. Thus, if one could show there exists a
product state in the ground-state space of two-body frustration-free
Hamiltonian containing $|\Psi\rangle$, the similar situation will
happen for any two-body frustration-free qubit Hamiltonian having
$|\Phi\rangle$ as a ground state. Therefore, we only need to
consider the SLOCC equivalent states with respect to the
transformation $L$.

We start from the simple case of three qubits. It is
known that there only two different SLOCC equivalent classes for
three-qubit genuinely entangled states~\cite{DVC00}, which are represented by
the $W$ state
\begin{equation}
|W\rangle=\frac{1}{\sqrt{3}}(|001\rangle+|010\rangle+|100\rangle),
\end{equation}
and the $GHZ$ state
\begin{equation}
|GHZ\rangle=\frac{1}{\sqrt{2}}(|000\rangle+|111\rangle).
\end{equation}
It is straightforward to check that
\begin{equation}
S(|W\rangle)=\text{span}\{|W\rangle,|000\rangle\};
\end{equation}
and
\begin{equation}
S(|GHZ\rangle)=\text{span}\{|000\rangle,|111\rangle\}.
\end{equation}
Apparently, both $S(|W\rangle)$ and $S(|GHZ\rangle)$ contain at
least one product state.

Next we proceed to the four-qubit case. Since $|\Psi\rangle$ is a
genuinely entangled state, the rank of every $\Pi_{ij}$ is at most
2.
%Let us denote the range of $\rho_{ij}$ by $\text{range}(\rho_{ij})$, and note that
%$\text{range}(\rho_{ij})=\text{ker}(\rho_{ij})^{\perp}$.
We then discuss two cases: all $\rho_{ij}$ are of rank $3$ or $4$,
or at least one $\rho_{ij}$ is of rank $2$.

If all $\rho_{ij}$ are of rank $3$ or $4$, then $H_{\Psi}$ is
actually the so-called homogenous Hamiltonian discussed
in~\cite{Bra06}. According to Lemma 2 of~\cite{Bra06}, one can
always find a product of single-qubit state in the ground-state
space of $H_{\Psi}$. We refer the readers to~\cite{Bra06} for
technical details on the discussion of homogenous Hamiltonians.

If for some pair of qubits, the rank of $\rho_{ij}$ is $2$. Without
loss of generality, we assume $\{i,j\}$ to be $\{3,4\}$. Now we can
reduce the problem to a three-qubit case by encoding qubits $3$ and
$4$ as one single qubit. Suppose $\rho_{34}$ is supported on two
orthogonal states $|\psi_0\rangle_{34}$ and $|\psi_1\rangle_{34}$.
Consider the following isometry,
\begin{equation}\label{eq:isometry}
V: |0\rangle_{3'}\rightarrow
|\psi_0\rangle_{34},|1\rangle_{3'}\rightarrow |\psi_1\rangle_{34},
\end{equation}
where qubits 3 and 4 are encoded by a new qubit $3'$.

Define
\begin{equation}
|\Phi\rangle=V^\dag|\Psi\rangle, \tilde{H}=V^\dag H_{\Psi}V.
\end{equation}
It can be seen that $\tilde{H}$ is also a two-body frustration-free
Hamiltonian, and $|\Phi\rangle$ is still genuinely entangled.
Besides, $|\Psi\rangle$ is a ground state of $H_{\Psi}$ if and only
if $|\Phi\rangle$ is a ground state of $\tilde{H}$.

Since $\tilde{H}$ is exactly the case of three qubits we have
already discussed, it follows immediately that the ground-state
space of $H'$ contains a product state
$|\alpha_1\rangle\otimes|\alpha_2\rangle\otimes|\alpha_{3'}\rangle$.
Let $V|\alpha_{3'}\rangle=|\beta_{34}\rangle$. If
$|\beta_{34}\rangle$ is product state, then $|\Psi\rangle$ is also a
product state. For the case that $|\beta_{34}\rangle$ is entangled,
one can always find a product state
$|\beta_1\rangle\otimes|\beta_2\rangle$ in the range of $\rho_{34}$,
since this is a two-dimensional subspace of two qubits~\cite{Par04}.

The case of $n>4$ qubits can just be similarly shown by induction.
Thus we finally conclude that for any entangled state $\ket{\Psi}$,
there always exists a product state of single qubits which is also a
ground state of $H_{\Psi}$. This also indicates that if we do not
require that $|\Psi\rangle$ being genuinely entangled, then the
ground-state space of $H_{\Psi}$ always contains a state which is a
product of single- or two-qubit states.

Therefore, we can conclude that any genuinely entangled $n$-qubit
state $\ket{\Psi}$ cannot be a unique ground state of a two-body
frustration-free Hamiltonian. Because that a resource state for MBQC
must be genuinely entangled, Conditions $1-5$ given in
Sec.~\ref{sec:resource} cannot be simultaneously satisfied for any
qubit system. In other words, there does not exist any idea resource
state for MBQC in spin-$1/2$ systems.

\begin{figure}[htbp]
  \centering
  \includegraphics{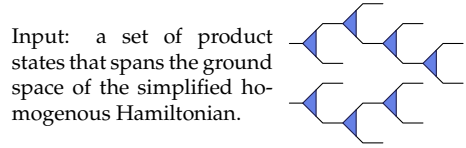}
  \caption{(FIG.1 in ~\cite{JWZ10}.) The general structure of the ground space of 2-body frustration-free qubit Hamiltonian}
  \label{fig:forest}
\end{figure}

It should be pointed out that the above discussion regarding the
ground-state space properties of two-body frustration-free
Hamiltonians of qubits can be strengthened further. In fact, it is
shown by Ji {\it et al.} that the structure of the ground-state
space of this kind of Hamiltonians can be characterized
completely~\cite{JWZ10}. That is, such a ground-state space is a
span of tree tensor network states of the same tree structure, which
is illustrated in Fig.~\ref{fig:forest} (we refer the readers
to~\cite{SDV06} for detailed discussion tree tensor network states).
Here we briefly discuss the structure given by Fig.~\ref{fig:forest}
as follows. For any two-body frustration-free Hamiltonian $H$ of
qubits, one can produce a corresponding special Hamiltonian $H'$,
called the simplified homogenous Hamiltonian (again, we refer the
readers to~\cite{Bra06} for technical details on the discussion of
simplified homogenous Hamiltonians). It can be shown that the
ground-state spaces of these simplified homogenous Hamiltonians are
actually spanned by (not necessarily orthogonal) product states of
single qubits. In Fig.~\ref{fig:forest}, every blue triangle stands
for an isometry defined in Eq.~\ref{eq:isometry}. The very left side
of the tree structure is the input, and it is the set of the product
states that span the ground-state space of the corresponding
simplified homogenous Hamiltonian $H'$. Then every product state
solution goes through the tree structure from the left to the right,
and after being operated by the meeting blue triangles, each of them
becomes a solution of the original Hamiltonian $H$. Finally, the
entire ground-state space of $H$ is then spanned by these tree
tensor network states.

\section{Future directions}

We have reviewed the recent results toward finding practical
resource states for MBQC. Starting from Verstraete and Cirac's
formulation in 2004~\cite{VC04} relating cluster states to VBS
states, we discuss a general method for producing VBS resource
states for MBQC by Gross and Eisert~\cite{GE07}. There has been good
progress made to identify practical resource states for MBQC with
VBS states. One-dimensional resource states that are useful for
single-qubit MBQC are identified for spin-$1$ systems, as well as
two-dimensional resource states for spin-$5/2$ and then later
spin-$3/2$ systems.

Down the road there are still many interesting open threads. The
no-go result by Chen {\it et al.} discussed in
Sec.~\ref{sec:spinonehalf} rules out the existence of any spin-$1/2$
system as a practical resource state for realizing MBQC, satisfying
all of Conditions $1-5$ given in Sec.~\ref{sec:resource}. As the
resource states satisfying all of Conditions $1-5$ are identified in
Sec.~\ref{sec:spinthreehalf} only for spin-$3/2$ systems, this
naturally leaves a gap on whether a resource state satisfying all of
Conditions $1-5$ can be identified in a spin-$1$ system. In other
words, it is still unknown whether there exists a spin-$1$ state on
some two-dimensional lattice which could be universal for MBQC, and
at the same time is the unique ground state of some frustration-free
and gapped Hamiltonian involving only two-body nearest-neighbor
interactions. So far such a state has only been found for
one-dimensional lattices, which does not rule out the existence of
such a state on a two-dimensional lattice.

Given that the spin-$1/2$ systems are the most commonly found
systems in nature, one would also like to seek for some relaxation
of the ideal Conditions $1-5$ for the systems. For instance, one may
try to relax the ``frustration-free'' condition (Condition $5$) or
the ``uniqueness'' condition (Condition $4$) requirement by
resorting to some other physical mechanisms, such as topological
protection as discussed in~\cite{Miy11,BBM+11} or perturbation as
discussed in~\cite{NLD+08,BR08,BO08}.

In a more general theoretical framework, each one of Conditions
$1-5$ lacks a deeper understanding. That is why the construction of
each resource state discussed in this review turns out be somewhat
``\textit{ad hoc}'', apart from the fact that they are all VBS
states.

For Condition $1$, we would like a general understanding concerning
which kinds of states could be universal for MBQC. In particular,
whether the observation discussed in Sec.~\ref{sec:resource}, that
is, if a spin state $\ket{\psi}$ can be reduced to a resource
cluster state $\ket{\Psi_C}$ via adaptive local measurements at a
constant cost, then $\ket{\psi}$ is a resource state for MBQC, is in
general a necessary condition for all resource states. So far, all
the examples of resource states discussed in this review satisfy
this condition, and indeed we use this condition to prove the
universality property for MBQC for all these states. It then remains
open whether a state which can be used for universal MBQC, but does
not satisfy this condition, can be identified.

For Conditions $2$ and $3$, we would like a general understanding
that which kinds of quantum states can be unique ground states of
two-body Hamiltonians of certain interaction patterns (for instance,
two-body nearest-neighbor interactions on some kind of lattices).
Recently, this problem has been linked to the certain kind of
correlations of quantum states by Chen et al. in~\cite{CJZ+11}. The
work along the line of Chen {\it et al.} raises a general question
of the so-called ``from ground states to local Hamiltonians''. That
is, to develop general understanding and method to find some desired
local Hamiltonians to have a given quantum state as its unique
ground state.

However, these local Hamiltonians constructed by Chen {\it et al.}
in~\cite{CJZ+11} are in general frustrated, and their discussion is
restricted to finite systems (i.e., systems composed of finite $n$
particles), so whether these Hamiltonians are gapped, which is
Condition $4$, needs further discussion.  It remains a general
challenge to determine whether a two-body Hamiltonian is gapped, and
one would like to obtain concrete examples showing whether the 2D
AKLT state on the honeycomb lattice is gapped, as mentioned in
Sec.~\ref{sec:AKLT2D}.

Finally, we would like to mention that Condition $5$, i.e., the
study of ground states of frustration-free Hamiltonians, is closely
related to computer science. Indeed, some of the techniques used in
Sec.~\ref{sec:spinonehalf}, have been used to deal with some quantum
version of a computer science problem in~\cite{Bra06}. This problem,
called the quantum $2$-satisfiability (2-SAT) problem, is of vital
importance in the study of the theory of quantum computational
complexity. Indeed, the characterization of the general structure of
the ground-state spaces of two-body frustration-free qubit
Hamiltonians shown by Fig.~\ref{fig:forest} has an immediate and
interesting corollary regarding quantum computational complexity
theory, i.e., the corresponding counting problem of quantum 2-SAT is
equal to its classical counterpart in computational complexity,
which answers an open problem raised in~\cite{BMR09}. More details
can be found in~\cite{JWZ10}.

We therefore believe that these recently progresses related to MBQC
with VBS states discussed in this review will not only push the
effort of realizing large scale quantum computers, but also open up
new directions that can further enhance the research in both quantum
information science and condensed matter physics.

\section*{Acknowlegement} We thank Xie Chen and Zhengfeng Ji for helpful discussions.
This work was supported in part by National Research Foundation \&
Ministry of Education, Singapore (L.-C. K. and Z.W.). Z.W. would
also like to acknowledge the WBS grant under contract no.
R-710-000-007-271. B.Z. is supported by NSERC and CIFAR.

\bibliography{ReviewForVBS}

\end{document}